\documentclass{emulateapj}

\usepackage{lscape}

\newcommand{\citepeg}[1]{\citep[{e.g.,}][]{#1}}

\shorttitle{Host Galaxies of PTF SLSNe}
\shortauthors{Perley et al.}

\begin{document}

\title{Host-Galaxy Properties of 32 Low-Redshift Superluminous Supernovae \\from the Palomar Transient Factory}

\def\dark{1}
\def\cit{2}
\def\sdsu{3}
\def\ipmu{4}
\def\ipac{5}
\def\wis{6}
\def\eso{7}
\def\ucb{8}
\def\ssc{9}
\def\lbnl{10}
\def\mail{*}

\author{D.~A.~Perley\altaffilmark{\dark,\cit,\mail},
R.~M.~Quimby\altaffilmark{\sdsu,\ipmu},
L.~Yan\altaffilmark{\ipac},
P.~M.~Vreeswijk\altaffilmark{\wis},
A.~De~Cia\altaffilmark{\wis,\eso}, 
R.~Lunnan\altaffilmark{\cit},
A.~Gal-Yam\altaffilmark{\wis}, \\
O.~Yaron\altaffilmark{\wis},
A.~V.~Filippenko\altaffilmark{\ucb},
M.~L.~Graham\altaffilmark{\ucb},
R.~Laher\altaffilmark{\ssc}, 
and
P.~E.~Nugent\altaffilmark{\ucb,\lbnl}}

\altaffiltext{\dark}{Dark Cosmology Centre, Niels Bohr Institute, University of Copenhagen, Juliane Maries Vej 30, 2100 K{\o}benhavn {\O}, Denmark}
\altaffiltext{\cit}{Department of Astronomy, California Institute of Technology, MC 249-17, 1200 East California Blvd., Pasadena CA 91125, USA}
\altaffiltext{\sdsu}{Department of Astronomy, San Diego State University, San Diego, CA 92182, USA}
\altaffiltext{\ipmu}{Kavli IPMU (WPI), UTIAS, The University of Tokyo, Kashiwa, Chiba 277-8583, Japan}
\altaffiltext{\ipac}{Infrared Processing and Analysis Center, California Institute of Technology, Pasadena, CA 91125, USA}
\altaffiltext{\wis}{Department of Particle Physics and Astrophysics, Faculty of Physics, The Weizmann Institute of Science, Rehovot 76100, Israel}
\altaffiltext{\eso}{European Southern Observatory, Karl-Schwarzschild-Strasse 2, 85748 Garching bei M\"unchen, Germany}
\altaffiltext{\ucb}{Department of Astronomy, University of California, Berkeley, CA 94720-3411, USA}
\altaffiltext{\ssc}{Spitzer Science Center, California Institute of Technology,  MC 314-6, Pasadena, CA 91125, USA}
\altaffiltext{\lbnl}{Lawrence Berkeley National Laboratory, 1 Cyclotron Rd., Berkeley, CA 94720, USA}
\altaffiltext{\mail}{e-mail: dperley@dark-cosmology.dk }
\slugcomment{Accepted to ApJ}

\begin{abstract}
We present ultraviolet through near-infrared photometry and spectroscopy of the host galaxies of all superluminous supernovae (SLSNe) discovered by the Palomar Transient Factory prior to 2013, and derive measurements of their luminosities, star-formation rates, stellar masses, and gas-phase metallicities.   We find that Type~I (hydrogen-poor) SLSNe are found almost exclusively in low-mass ($M_*<2\times10^9\,{\rm M}_\odot$) and metal-poor (12 + log$_{10}$[O/H] $< 8.4$) galaxies.  We compare the mass and metallicity distributions of our sample to nearby galaxy catalogs in detail and conclude that the rate of SLSNe-I as a fraction of all SNe is heavily suppressed in galaxies with metallicities $\gtrsim 0.5\,{\rm Z}_\odot$.  Extremely low metallicities are not required, and indeed provide no further increase in the relative SLSN rate.  Several SLSN-I hosts are undergoing vigorous starbursts, but this may simply be a side effect of metallicity dependence: dwarf galaxies tend to have bursty star-formation histories.   Type~II (hydrogen-rich) SLSNe are found over the entire range of galaxy masses and metallicities, and their integrated properties do not suggest a strong preference for (or against) low-mass/low-metallicity galaxies.  Two hosts exhibit unusual properties: PTF\,10uhf is a Type I SLSN in a massive, luminous infrared galaxy at redshift $z=0.29$, while PTF\,10tpz is a Type II SLSN located in the nucleus of an early-type host at $z=0.04$.
\end{abstract}


\section{Introduction}
\label{sec:intro}

The recently discovered observational class of ``superluminous'' supernovae (SLSNe) has complicated what was once a fairly straightforward view of the fates of massive stars in the local universe in which all stars above $8\,{\rm M}_\odot$ were thought to explode via a common mechanism of iron core collapse \citep[see, e.g.,][for a review]{Filippenko1997}.  SLSNe have characteristic peak visual absolute magnitudes between $-21$ and $-22.5$ ($\sim10^{10}\,{\rm L}_{\odot}$; \citealt{GalYam2012,Nicholl+2015}), making them much more luminous than typical core-collapse supernovae (CC-SNe), which peak between $-15$ and $-18$\,mag ($\sim10^8-10^9\,{\rm L}_{\odot}$; \citealt{Richardson+2002,Li+2011}.)  Most SLSNe also evolve much more slowly and have higher peak temperatures than ordinary CC-SNe, and the time-integrated bolometric radiative output of a SLSN may reach $\gtrsim10^{51}$ erg, exceeding a typical CC-SN by 2--3 orders of magnitude.  This points to a much larger progenitor mass, and may require a fundamentally different explosion mechanism.

The spectroscopic properties of SLSNe are diverse: they include events showing strong hydrogen emission throughout their observed evolution, events that show no hydrogen lines at any epoch, and intermediate cases of weak and/or transient hydrogen emission.  Mirroring the classification scheme for ordinary SNe, SLSNe are classified as Type I (no hydrogen observed) or Type II (hydrogen observed); see \citealt{GalYam2012} for a review of SLSN classes and properties. 

Events showing narrow or intermediate-width hydrogen lines in their spectra (all of which are Type II by definition, and which represent the majority of events in this class) are simplest to accommodate physically, since the existence of these lines is direct evidence of interaction between SN ejecta and a dense surrounding medium \citepeg{Moriya+2013}.  This process permits the bulk kinetic energy of the outflow to be tapped and converted to electromagnetic radiation, helping to explain the large radiative output of these events and easing the fundamental energy requirements.  Indeed, ``ordinary'' Type IIn SNe are the most luminous class of CC-SNe and are thought to occur when SN ejecta collide with shells of material from previous eruptions \citepeg{Schlegel+1990,Filippenko1997,Kiewe+2012}.  The underlying mechanism in SLSNe-IIn is presumably directly analogous.  Nevertheless, the amount of kinetic energy that must be converted to radiation in order to accommodate these events requires an extremely massive circumstellar envelope and therefore a very large initial mass \citep{Smith+2010,Chatzopoulos+2013}, possibly within the range at which other evolutionary channels beyond ordinary core collapse may become relevant.  

Events lacking narrow hydrogen lines (including all SLSNe-I, but also any SLSNe-II whose Balmer lines are broad and/or weak) represent an even greater challenge for progenitor models, since it is not clear whether interaction with a circumstellar medium is available to ease the radiative energy requirements.  It is possible that the ejecta are interacting with a dense, hydrogen-poor shell of previously ejected material \citepeg{Chevalier+2011,Ginzburg+2012}, although this would imply a very large initial mass and again point toward the possibility of exotic evolutionary or explosion channels.  Furthermore, the absence of observed narrow lines from other elements is surprising.  For SLSNe-I, the stellar progenitor must also rid itself of its hydrogen envelope during its lifetime and yet retain sufficient mass at the time of death to produce an explosion with $E_K\approx10^{51}$ ergs ejecting $M_{\rm ej}\gtrsim10\,{\rm M}_\odot$ worth of heavy elements, a challenge for stellar evolutionary theory.

While it is possible that either or both classes of transient may simply constitute extrema of ordinary stellar-evolution and explode via core collapse, the remarkable observational properties of SLSNe have sparked renewed interest in more exotic explosion mechanisms.
One well-established model of particular theoretical interest is the pair-instability supernova (PI-SN), an explosion produced when the temperature required to maintain hydrostatic equilibrium in the core of a star becomes so high that photons disintegrate into particle pairs and the star collapses \citep{Rakavy+1967,Barkat+1967}.  Such events should originate from the most massive stars ($M_{\rm init} > 300\, {\rm M}_\odot$; \citealt{Yoshida+2011}) and are expected to produce enormous quantities of radioactive nickel that could easily power a SLSN; at least one well-known SLSN-I (SN\,2007bi) has been interpreted with reasonable success within this model (\citealt{GalYam+2009}; but see \citealt{Dessart+2012}).   In a variant on the pair-instability mechanism, the \emph{pulsational} pair-instability supernova, the massive star undergoes several incomplete pair-instability episodes leading to a series of envelope-shedding eruptions before the final explosion \citep{Woosley+2007,Waldman+2008}, naturally providing both an intrinsically very energetic explosion and shells of material for it to interact with.  Problematically, however, classical pair-instability models lead to very large masses of $^{56}$Ni; this decays to $^{56}$Co, whose much slower decay to $^{56}$Fe should produce a luminous exponential decay phase in the late-time light curve.  While evolution consistent with this has been seen in a handful of cases (referred to as ``Type R'' SLSNe by \citealt{GalYam2012}), the majority fade too fast to be explained by this mechanism.  High-mass, noninteracting core-collapse models \citepeg{Yoshida+2014} also share this problem.

Some, and perhaps all, SLSNe may therefore require yet another mechanism to re-energize the ejecta.  If interaction and radioactive decay are excluded as possibilities, the only remaining power source capable of meeting the energy requirements is the compact object itself, the so-called ``central engine''.  The most popular central-engine model invokes a spinning-down highly magnetic neutron star (``magnetar'') that energizes the SN by winds and X-ray radiation from inside \citep{Mazzali+2006,Kasen+2010,Woosley2010,Inserra+2013,Metzger+2015}.  Alternative central-engine models include jet feedback from fallback accretion onto the central neutron star or black hole \citep{Gilkis+2015,Soker2016}.

The host-galaxy environments of SLSNe provide strong constraints on progenitor models.  For example, simple, single-star pair-instability models predict that pair-instability SNe should be produced only by stars with very low initial metallicity (\citealt{Langer+2007}).  
If this model is correct, these explosions should not form in metal-rich environments.  The energy-injection model involves a rapidly rotating central engine similar in nature to the central engine of gamma-ray bursts \citep[GRBs; e.g.,][]{Usov+1992}; if this model explains some or all SLSNe then it would be reasonable to expect similarities between the hosts of SLSNe and the hosts of long-duration GRBs, which are observed to avoid high-metallicity galaxies and occur predominantly at low-to-intermediate metallicity in the local universe \citep{Stanek+2006,Modjaz+2008,Graham+2013,Kruehler+2015,Perley+2016b,Japelj+2016}.  Other models invoke dynamical interactions and stellar mergers in dense environments \citep{Pan+2012,vandenHeuvel+2013}, which would favor particularly intense starbursts. In any case, regardless of the underlying theoretical model, the degree of similarity or dissimilarity between the hosts of Type I versus Type II events (or between subclasses of these events, or between these events and other classes of SNe) might help establish whether these explosions are closely related or fundamentally different.

\begin{deluxetable*}{llll llll l}  
\tabletypesize{\small}
\tablecaption{Superluminous Supernovae from PTF}
\tablecolumns{9}
\tablehead{
\colhead{PTF ID} &
\colhead{$\alpha$\tablenotemark{a}} & 
\colhead{$\delta$\tablenotemark{a}} &
\colhead{class\tablenotemark{b}} &
\colhead{$z$} & 
\colhead{$M_V,_{\rm peak}$\tablenotemark{c}} & 
\colhead{$t_{\rm peak}$\tablenotemark{c}} &
\colhead{$E_{B-V}$\tablenotemark{d}} &
\colhead{Notes}
}
\startdata
09as   & 12:59:15.864 & $+$27:16:40.58 & I    & 0.1867 &$-20.8 $& 2009-03-24  & 0.008 &  \\
09uy   & 12:43:55.771 & $+$74:41:07.58 & II   & 0.3145 &$-21.2 $& 2009-07-08  & 0.020 &  \\
09atu  & 16:30:24.553 & $+$23:38:25.43 & I    & 0.5015 &$-22.5 $& 2009-08-18  & 0.042 &  \\
09cnd  & 16:12:08.839 & $+$51:29:16.01 & I    & 0.2584 &$-23   $& 2009-09-10  & 0.021 &  \\
09cwl  & 14:49:10.08  & $+$29:25:11.4  & I    & 0.3499 &$-22.5 $& 2009-08-07  & 0.014 & = SN\,2009jh \\
10bfz  & 12:54:41.288 & $+$15:24:17.08 & I    & 0.1701 &$-20.9 $& 2010-01-31  & 0.018 &  \\
10bjp  & 10:06:34.30  & $+$67:59:19.0  & I    & 0.3584 &$-21.4 $& 2010-02-16  & 0.055 &  \\
10cwr  & 11:25:46.73  & $-$08:49:41.9  & I    & 0.2297 &$-21.8 $& 2010-03-21  & 0.035 & = SN\,2010gx \\
10fel  & 16:27:31.103 & $+$51:21:43.45 & II   & 0.2356 &$<-20.5$&$<$2010-04-03& 0.017 &  \\
10heh  & 12:48:52.05  & $+$13:26:24.5  & II   & 0.3379 &$-21.2 $& 2010-06-03  & 0.024 &  \\
10hgi  & 16:37:47.074 & $+$06:12:31.83 & I    & 0.0987 &$-20.3 $& 2010-06-20  & 0.074 & = SN\,2010md \\
10jwd  & 16:43:43.325 & $+$44:31:43.8  & II   & 0.477  &$-21.4 $& 2010-07-02  & 0.012 &  \\
10nmn  & 15:50:02.809 & $-$07:24:42.38 & I-R  & 0.1237 &$-20.5 $& 2010-07-07  & 0.138 &  \\
10qaf  & 23:35:42.887 & $+$10:46:32.57 & II   & 0.2836 &$-21.6 $& 2010-08-05  & 0.070 &  \\
10qwu  & 16:51:10.572 & $+$28:18:07.62 & II   & 0.2258 &$-21.0 $& 2010-08-21  & 0.040 &  \\
10scc  & 23:28:10.495 & $+$28:38:31.10 & II   & 0.242  &$-21.5 $& 2010-08-26  & 0.093 &  \\
10tpz  & 21:58:31.74  & $-$15:33:02.6  & II   & 0.0395 &$<<-19 $& 2010-09-02  & 0.041 & Heavily extinguished \\
10uhf  & 16:52:46.696 & $+$47:36:21.76 & I    & 0.2882 &$-22   $& 2010-09-18  & 0.018 & Possible very weak H$\alpha$? \\
10vqv  & 03:03:06.859 & $-$01:32:35.42 & I    & 0.4518 &$-22.5 $& 2010-10-13  & 0.061 &  \\
10vwg  & 18:59:32.881 & $+$19:24:25.74 & I-R  & 0.1901 &$-21   $& 2010-09-07  & 0.467 & = SN\,2010hy.  KAIT/LOSS discovery. \\
10yyc  & 04:39:17.297 & $-$00:20:54.5  & II   & 0.2147 &$-21   $& 2010-11-13  & 0.041 &  \\
10aagc & 09:39:56.923 & $+$21:43:17.09 & I    & 0.206  &$-20.4 $& 2010-10-04  & 0.022 & Late-time hydrogen lines? \\
11dij  & 13:50:57.798 & $+$26:16:42.44 & I    & 0.1428 &$-21.5 $& 2011-04-28  & 0.011 & = SN\,2011ke \\
11dsf  & 16:11:33.55  & $+$40:18:03.5  & II   & 0.3848 &$-22.1 $& 2011-05-27  & 0.009 &  \\
11hrq  & 00:51:47.22  & $-$26:25:10.0  & I    & 0.057  &$<-20  $&$<$2011-07-11& 0.012 &  \\
11rks  & 01:39:45.528 & $+$29:55:27.43 & I    & 0.1924 &$-21.1 $& 2012-01-11  & 0.038 &  \\
12dam  & 14:24:46.228 & $+$46:13:48.64 & I-R  & 0.1073 &$-21.5 $& 2012-06-12  & 0.100 &  \\
12epg  & 12:55:36.596 & $+$35:37:35.79 & II   & 0.3422 &$-21.3 $& 2012-05-30  & 0.015 &  \\
12gwu  & 15:02:32.876 & $+$08:03:49.47 & II   & 0.275  &$-21.4 $& 2012-07-25  & 0.033 & Hydrogen lines very weak. \\
12mkp  & 08:28:35.092 & $+$65:10:55.60 & II   & 0.153  &$-21.0 $& 2013-01-25  & 0.046 &  \\
12mue  & 03:18:51.072 & $-$11:49:13.55 & II   & 0.2787 &$-21.4 $& 2012-12-21  & 0.062 &  \\
12mxx  & 22:30:16.728 & $+$27:58:22.01 & I    & 0.3296 &$-22.5 $& 2012-12-10  & 0.041 &  \\
\enddata
\label{tab:slsne}
\tablenotetext{a}{Supernova position (J2000)}
\tablenotetext{b}{Supernova classification}
\tablenotetext{c}{Approximate peak visual magnitude of the supernova and corresponding UT date.  More refined measurements will be presented by De Cia et al., in prep.}
\tablenotetext{d}{Galactic (foreground) selective extinction in magnitudes; from \cite{Schlafly+2011}.}
\end{deluxetable*}

The very fact that SLSNe were discovered only in the past decade provides evidence that the sites of SLSNe might differ from those of ordinary CC-SNe.  Prior to about 2005, all major nearby SN searches---most notably, the Lick Observatory Supernova Search (LOSS) with the Katzman Automatic Imaging Telescope (KAIT; \citealt{Filippenko+2001})---were targeted surveys, using small-field-of-view cameras to periodically image the positions of known galaxies.  For reasons of efficiency, nearby and relatively high-mass galaxies were preferentially targeted, rendering these searches insensitive to transients that might occur preferentially or exclusively in smaller systems (unless discovered in the background).   However, starting about 10\,yr ago a number of wide-field untargeted optical surveys began operation, providing the capability to search much larger volumes of space in an unbiased manner; these include the Texas Supernova Search (which discovered the first widely recognized SLSNe, SN\,2005ap and SN\,2006gy), the Catalina Real-Time Survey \citep{Drake+2009}, the Palomar Transient Factory \citep{Law+2009}, Pan-STARRS \citep{Kaiser+2002}, La Silla Quest \citep{Hadjiyska+2012}, SkyMapper \citep{Keller+2007}, the Dark Energy Survey \citep{DES+2016}, and the All-Sky Automated Survey for Supernovae (ASAS-SN; \citealt{Shappee+2014}).  A large fraction of the SLSNe reported by these surveys originate from very faint galaxies \citep{Neill+2011}, undetected in pre-explosion images such as the Sloan Digital Sky Survey (SDSS).  While in part this reflects the great distances at which SLSNe are discovered, more detailed analysis of SLSN host-galaxy samples suggests that they differ intrinsically from the host populations of more ordinary SNe in various ways: low masses and metallicities are typical \citep{Chen+2013,Lunnan+2013,Lunnan+2014,Angus+2016}, and galaxies with exceptionally strong emission lines are remarkably frequent \citep{Leloudas+2015}.

Among these surveys, the Palomar Transient Factory (PTF) has been the most prolific discoverer of SLSNe: the sample of 32 SLSNe discovered in 2009--2012 that we present here (\S \ref{sec:sample}) is comparable in size to the sample of publicly released SLSNe from all other surveys combined.   Furthermore, all of these events occurred at relatively low redshifts ($z<0.51$), so in all cases the SN and host are relatively accessible to comprehensive follow-up observations.  In contrast, Pan-STARRS, the next most-prolific individual survey with a published SLSN sample, has discovered most of its SLSNe at significantly greater distances ($0.5 < z < 1.6$ from the sample of \citealt{Lunnan+2014}).

In complementary papers we will be presenting the entire suite of observations of the PTF SLSN sample, including details of the discovery and sample selection, spectroscopic properties (Quimby\ et al. 2016; Leloudas et al. 2016), as well as multiband light curves (De Cia et al. 2016).  In this work we present observations of the host galaxies of these events from an extensive ground- and space-based campaign, effectively doubling the sample of well-studied SLSN hosts and providing the first large, homogeneous, single-survey sample in the local universe.  

The paper is organized as follows.  In \S \ref{sec:sample} we overview the operations in PTF leading to successful discovery and classification of SLSNe and outline the selection of our sample.   Our observations are described in \S\ref{sec:observations}, including ultraviolet (UV), optical, and near-infrared (NIR) photometry and spectroscopy from Keck and Palomar supplemented by {\it Hubble Space Telescope (HST)} and {\it Spitzer} imaging; we also summarize our analysis techniques used to provide measurements of physical parameters such as mass, star-formation rate (SFR), and metallicity using these observations.  The host galaxies are discussed on an individual basis in \S\ref{sec:individuals}.   In \S\ref{sec:results} we examine our SLSN sample as an ensemble and compare the physical properties of the population against those of volume-limited star-forming field-galaxy samples.  We discuss our results and their implications in \S\ref{sec:discussion} and \S\ref{sec:conclusions}.

\vskip 0.7cm  

\section{Sample Selection}
\label{sec:sample}

\subsection{PTF Discovery of Supernovae}

The current public literature sample of SLSNe and SLSN hosts (see, e.g., \citealt{Leloudas+2015,Angus+2016}) is combined from a large variety of different surveys, each of which contributes only a few events to the overall total.\footnote{The relatively large Pan-STARRS sample of 15 events presented by \citealt{McCrum+2015} and \citealt{Lunnan+2014} is an exception, but it probes a higher and more difficult-to-study redshift range.}  Many of these discoveries were based on archival re-analysis of earlier surveys (such as SDSS; \citealt{Leloudas+2012}) to recover events that were not recognized to be SLSNe at the time.  Consequently, the existing sample of low-$z$ SLSNe is quite heterogeneous in construction and the biases which may affect the nature of the catalogued population are nontrivial.  
In contrast, the PTF sample we present here was discovered by a single survey using a single camera and telescope and via (typically) the same group of scientists and follow-up resources.  Nevertheless, PTF is a complex effort, and its cadence, motivations, and emphasis have varied substantially since its inception, so the sample presented here is subject to its own biases and incompletenesses.  Discussion of possible biases related to these factors will be presented in \S \ref{sec:selectioneffects}.  We provide a brief summary of the survey and its operations below.

The Palomar Transient Factory is a synoptic optical survey using the 48-inch Oschin Schmidt Telescope (P48) at the Palomar Observatory near San Diego, California and a 7.2 deg$^2$ camera \citep{Rahmer+2008}.  Observations of the sky are acquired every night during clear weather, except within a few days of full moon each month when H$\alpha$ survey observations are performed.  PTF operated between 2009 and 2012, and although the facility is continuing operations as the intermediate Palomar Transient Factory (iPTF) until the end of 2016, this paper exclusively addresses events discovered during the original four-year period.  PTF employs both $R$- and $g$-band filters, but prior to 2013 the large majority of the survey was conducted in $R$, and all of the SLSNe presented here were discovered in $R$.

The survey discovers far more transient events than can be observed spectroscopically: the PTF database reports 19595 likely transients discovered in 2009--2012, of which only 2131 (11\%) have secure classifications.  Human oversight is necessary at several stages in the process to choose astrophysically real and scientifically interesting targets for follow-up observations.   All objects found by the automated detection and verification pipelines (\citealt{Brink+2013}) are screened by human scanners to confirm their astrophysical nature and rule out nontransient false positives (cosmic rays, poor subtractions, asteroids, and variable sources).  At the time of discovery, the scanner may choose to nominate an object for follow-up spectroscopy.  Objects may also be nominated later, as further data are collected.  Weather permitting, these are then targeted at the next observing run (usually 1--2 runs occur monthly during dark time).   Spectra are reduced within a few days of being acquired and a preliminary classification is established either visually or via standard classification routines; events with unclear or ambiguous classifications are flagged for re-observation with higher signal-to-noise ratio (S/N) or at later epochs.  Classifications are revisited at later times once all data are in hand.

\subsection{Definition and Identification of SLSNe}
\label{sec:classification}

The class of SLSNe is necessarily defined via a combination of photometric and spectroscopic qualities: to qualify, an event must be clearly a SN (usually implying the detection of broad features in the spectrum, as well as an SN-like light curve and color evolution) and also must be much more luminous than ordinary SNe (it must be ``super'' luminous).  Beyond this there is no standard definition of what observables are required to establish what is or is not a SLSN. 
An absolute magnitude limit of $M <-21$ at peak was adopted by \cite{GalYam2012}, but this choice is empirical and somewhat arbitrary (it is also wavelength-dependent).  Furthermore, several SNe with properties very similar to those of Type I SLSNe in particular (in terms of colors, light curves, spectra, and total radiative output) do not quite reach this luminosity, while a small number of SNe that are probably not related to massive stars at all (specifically the Type Ia-CSM SNe; \citealt{Silverman+2013}) occasionally do surpass it.

The task of defining SLSNe in a physically meaningful way, and the isolation of all events within PTF satisfying that definition, is therefore quite complicated.  A detailed analysis of this topic will be deferred to the upcoming dedicated works of Quimby et al., De Cia et al., and Leloudas et al., including a presentation of all spectra and light curves.  For the purposes of this paper, we establish our own working definition of SLSNe in the PTF sample as follows.

We require, at minimum, an absolute magnitude of $M_R <-20.0$ at peak to consider inclusion of an event in our sample.  This guarantees that every event in our sample is indeed very luminous and eliminates the vast majority of ordinary SNe in the PTF sample.  Circumstellar interaction is capable of significantly boosting the luminosity of all types of SNe \citep{Ofek+2014}; indeed, Type IIn SNe have in particular been known since the 1980s to exceed this threshold on occasion \citep{Richardson+2002}.  We therefore apply a more stringent cut if narrow hydrogen lines are present, requiring $M_R < -20.5$.   (A few events were discovered after peak brightness and one is heavily extinguished by host-galaxy dust.  In these cases peak magnitudes require extrapolations or corrections; see \S \ref{sec:sampleproperties}.)

Many of the most luminous transient candidates identified by PTF turn out to not be SNe: active galactic nuclei (AGNs) and quasi-stellar objects (QSOs) are particularly common.  Most such objects can be easily eliminated from consideration based on their past or continued variability or via spectroscopy; alternatively, an off-nuclear location or a smoothly rising and falling light curve with blue-to-red spectral evolution usually provides good evidence that a transient is an SN and not an AGN.  Even so, Type II SLSNe can look spectroscopically similar at certain phases to narrow-line AGNs (as can normal SNe~IIn; e.g., \citealt{Filippenko1989}), and in cases where photometric and spectroscopic coverage of the SN is poor it is not always easy to completely rule out an AGN flare.  For two events in our sample (PTF\,09uy and PTF\,11dsf), we favor a SLSN interpretation but note that an AGN has not been fully eliminated (see also the discussion of PTF\,10tpz in \S \ref{sec:10uhf10tpz}).  These classifications will be further investigated and discussed by Leloudas et al. (in prep.).

Tidal disruption events (TDEs) represent another, less-frequently observed class of phenomena associated with accretion onto supermassive black holes.  These typically exhibit peak magnitudes around $-19$ but can occasionally be brighter than $-20$ \citep{Arcavi+2014}.  The spectroscopic and photometric properties of TDEs and SLSNe are usually distinct---and while ambiguous cases can arise especially at the high-luminosity end (\citealt{Chornock+2014}; \citealt{Leloudas+2016}; Duggan et al.\ in prep.; see also the last paragraph of \S \ref{sec:conclusions}), these are particularly rare and we identify no such cases within the four-year PTF sample covered here.

The most luminous Type Ia SNe exceed $-20$ mag but are easily identified spectroscopically.  Type Ia-CSM SNe can be even brighter (occasionally, even $<-21$ mag), but as photospheric SN~Ia features are still evident these can similarly be identified spectroscopically \citep{Silverman+2013}. In the course of this analysis we identified several new SNe~Ia-CSM within the PTF sample that will be reported in separate work.

Other types of luminous transients are also known to exist whose connection to SNe is not yet clear, in particular the fast-rising transients of \cite{Arcavi+2016} and \cite{Drout+2014}.  With the exception of the single event already identified by \cite{Arcavi+2016}, we find no further members of these classes in our sample.

In total, 32 events satisfy all of the above criteria and constitute the PTF SLSN sample.  All show behavior characteristic of SNe, including broad spectral lines, evolutionary timescales of months, and blue-to-red spectral evolution in cases where multiband data are available.  

\subsection{Subclassification of SLSNe}
\label{sec:subclassification}

SLSNe within the sample are then subcategorized spectroscopically as Type ``I'' or ``II'' based  on the absence or presence (respectively) of hydrogen in their spectra.  While in principle this is a straightforward distinction, it conceals some complexity.  For example, a few SLSNe show no hydrogen in any of their early-time spectra but then develop broad hydrogen lines at late times \citep{Miller+2009,Gezari+2009,Benetti+2014,Yan+2015}.  Even for events that do exhibit hydrogen emission in all existing spectra, this emission can sometimes be relatively weak and/or exhibit no narrow component.
While a strictly literal interpretation would classify these events as Type II, some of them may quite plausibly be physically more closely related to Type I SNe (or represent an intermediate case or another class entirely; see also \citealt{Inserra+2016}).

In spite of these occasional ambiguities, the ``I'' versus ``II'' distinction is sufficient for the vast majority of events in our sample:  nearly all events without hydrogen at the time of discovery never show hydrogen in any follow-up spectra, and nearly all events with hydrogen exhibit strong emission lines at all phases including a narrow component.  The possible exceptions include PTF\,10aagc (Type I with ambiguous, weak, late-time broad hydrogen), PTF\,10uhf (Type I, but with a possible faint signature of broad Balmer emission that is difficult to disentangle from the host [N~II] emission), and PTF\,12gwu (type II, but the hydrogen lines are much weaker than in the rest of our sample and no obvious narrow component is present).  For this paper, we maintain the initial, conservative classifications of these events from the presence or absence of unambiguous hydrogen in their discovery spectra.

Among the Type I SLSNe, we denote a small number of events (three) as belonging to the subclass of long-lived ``R''-types, which show exponentially declining late-time light curves consistent with radioactive decay and which have been suggested to be examples of pair-instability SNe \citep{GalYam+2009,GalYam2012}, although this interpretation is contested by other authors \citepeg{Dessart+2012,Nicholl+2013,Jerkstrand+2016}.  We will generally refer to them as Type ``I-R.''  With only three events, we do not have sufficient sample size to perform a statistically robust comparison between the host properties of these events versus the more rapidly declining SNe~I, but as we observe no strong distinction between the host properties of these events and other Type I SLSNe in our sample (and other authors have reported similar results; e.g., \citealt{Lunnan+2014,Leloudas+2015}), we will generally treat all Type I SLSNe together in our analysis regardless of their light-curve properties.

\begin{figure*}
\centerline{
\includegraphics[width=16cm,angle=0]{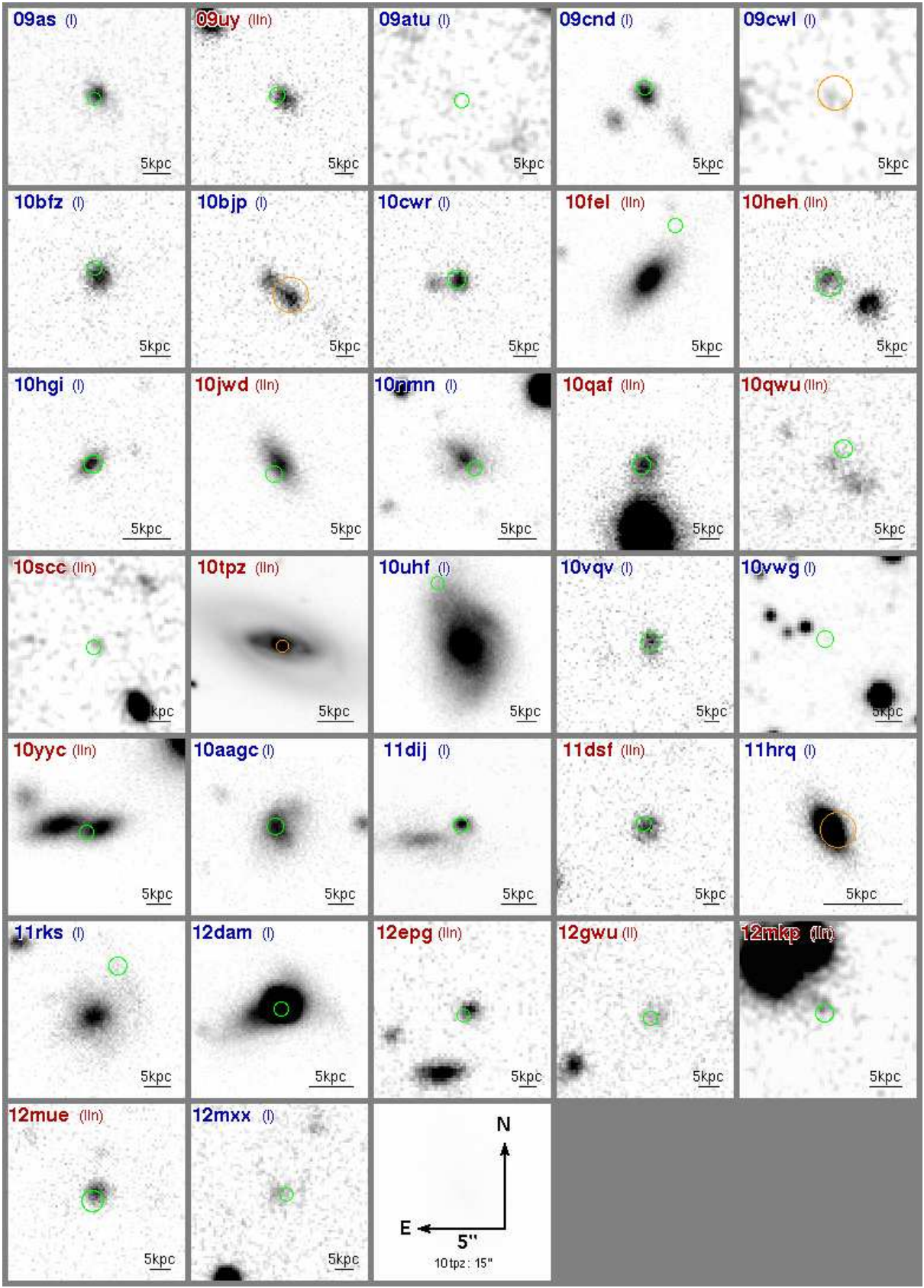}} 
\caption{Mosaic showing ground-based images of 32 PTF SLSN host galaxies from Keck and Palomar ($R$ or $I$-band images except for PTF\,10vwg and PTF\,12epg where we display $J$-band images, and PTF\,09as where we display a $u$-band image).  Images are 10$^{\prime\prime}$ on each side, except for the image of PTF\,10tpz (30$^{\prime\prime}$).  The host population exhibits a wide variety of morphologies, sizes, and luminosities, although mergers/companions are very common, and massive spirals are largely (but not entirely) deficient.   Circles show the SN position with the radius denoting the approximate uncertainty (green circles indicate positions from the P60 follow-up telescope, orange circles are from the P48 survey telescope).}
\label{fig:mosaic}
\end{figure*}

\subsection{Sample Properties}
\label{sec:sampleproperties}

The sample is summarized in Table \ref{tab:slsne}.  In total, we present 14 events of Type II and 18 events of Type I, only three of the latter being Type R.  The large majority of these objects have not been previously presented in the literature.

Two of the SLSNe are noteworthy from the point of view of sample selection.  PTF\,10vwg is in a crowded low-Galactic-latitude field and, while detected in PTF survey images, was not identified as a transient candidate until it was discovered in the background of a LOSS/KAIT image \citep{Kodros+2010}---so it is not truly a PTF object.  (Stellar confusion and high foreground extinction also introduce severe complications in characterizing the host.)  PTF\,10tpz is near the nucleus of an early-type galaxy and the SN spectrum is highly reddened owing to host extinction.  Without an extinction correction it would not be superluminous, but depending on the (highly uncertain) host column it is probably close to or above our threshold.  Were it not for the known example of SN\,2006gy, which occurred under similar circumstances \citepeg{Smith+2007,Ofek+2007}, this event would likely not have been categorized as a SLSN.  In any case, the considerations involving its discovery are quite different than for other PTF events (a much smaller effective detection volume because of the large extinction, plus more complex issues involving host subtraction and AGN contamination), so it should not be treated with statistical weight equal to the others.  We will include the hosts of these events in our plots and analysis where possible, but we emphasize that they would be excluded from any attempt to produce a statistically uniform sample from these data.

\section{Observations}
\label{sec:observations}

\subsection{Ground-Based Imaging}

The SLSNe in the sample above were targeted for late-time imaging in a variety of wavebands spanning the near-UV to the NIR.  Several host galaxies were bright enough to be well recovered in SDSS (at least in the $gri$ filters), so we downloaded the processed survey images from the SDSS archive \citep{Alam+2015}.  For fainter hosts and for other filters we observed with other facilities: the Palomar 60-inch (P60) telescope imaging camera \citep{Cenko+2006}, the Large Format Camera (LFC) or the Wide-Field Infrared Camera (WIRC) on the Palomar 5-m Hale telescope, or the Low-Resolution Imaging Spectrometer (LRIS; \citealt{Oke+1995}) or the Multi-Object Spectrometer for Infrared Exploration (MOSFIRE; \citealt{McLean+2012}) on the Keck I 10-m telescope.

All ground-based images used for the host-galaxy spectral energy distribution (SED) analysis were acquired either pre-explosion or long after the SN peak time (at least 2\,yr, typically 3--4\,yr).  Nevertheless, SLSNe exhibit a range of light-curve behaviors and decay times, and there is no guarantee based on the time difference alone that the SN is not contributing light.   We use various checks appropriate to the situation to rule out significant SN contributions: (1) direct confirmation based on a reference-subtracted image that the SN was much fainter than the host well prior to the observation in question, (2) verifying nonvariability between widely spaced epochs, (3) absence of any SN-like broad features in high-S/N contemporaneous (or earlier) spectra, or detection of features whose contribution to the overall flux is negligible, (4) a clearly resolved host with no point-like component visible at the SN location.

Observations were reduced via standard procedures (our own tools were developed and used for the reduction of LRIS, MOSFIRE, and WIRC data).  We localized the SN position by comparison to images taken with the P60 at early times while it was bright, though in a handful of cases no P60 imaging was available and we used the P48 survey images instead.  In nearly all cases we identify a host galaxy directly underlying this location at late times. For a few objects the nearest well-detected source is somewhat offset from the SN position or there is ambiguity regarding whether sources in the image constitute a single galaxy or multiple galaxies; we will discuss these cases individually in \S \ref{sec:individuals}.  

Once the host is identified, we measure its centroid and perform aperture photometry using our own IDL tools, setting the aperture radius for each galaxy to be sufficiently large as to include all of the host flux. The same aperture radius is used for all filters.  In one case (PTF\,11dij) a neighboring galaxy unavoidably contaminates the host position in all ground-based images; we modeled and subtracted it with \texttt{galfit} \citep{Peng+2002} before performing photometry on the subtracted images.  The photometric calibration scale is established either by direct comparison to SDSS/2MASS \citep{Cohen+2003}, or (for optical fields outside SDSS) from a field calibration of secondary standards gathered using the P60 on photometric nights.  We omit photometry that results only in shallow upper limits that do not usefully constrain the SED models.  We also omit a few points from the literature which are inconsistent with our own photometric measurements in the same or similar bands at high significance.

Thumbnail images of all our fields are presented in Figure \ref{fig:mosaic}.
Final photometry, including supplementary observations from the literature (\citealt{Lunnan+2015}; \citealt{Angus+2016}), is presented in Table \ref{tab:photometry} and displayed in Figure \ref{fig:seds}.  We report both magnitudes (in the default calibration system of the relevant filter and uncorrected for foreground extinction) and extinction-corrected fluxes (using the dust maps of \citealt{Schlafly+2011}).

\subsection{HST Observations}

Several host galaxies of slowly declining (Type I-R) hydrogen-poor SLSNe were observed with {\it HST} as part of our team's approved programs (GO-12983, PI O. Yaron; GO-13858, PI A. De Cia).  PTF\,10nmn was observed with the Advanced Camera for Surveys (ACS) in the Wide Field Channel through the F625W filter, while PTF\,12dam and PTF\,11hrq were observed with the Wide Field Camera 3 (WFC3) in the Ultraviolet-Visible (UVIS) channel using filters F625W, F336W, and F225W.

For the ACS observations, reduced images of the separate exposures were produced by the ACS data calibration pipeline (\texttt{CALACS}), which includes a correction for the bias striping and crosstalk effects and the charge transfer efficiency (CTE). For the UVIS observations, we corrected for the CTE using the \texttt{ctereverse} FORTRAN routine provided\footnote{\url{http://www.stsci.edu/hst/wfc3/tools/cte\_tools}} by the Space Telescope Science Institute (STScI). After cosmic ray removal using the LA Cosmic routine of \cite{vanDokkum+2001}, we astrodrizzled the reduced frames to a final image with DrizzlePac 2.0\footnote{\url{http://drizzlepac.stsci.edu}} with inverse variance map (IVM) weighting, adopting a pixel size of 0.033$\arcsec$.  If only two images were available (for the PTF\,11hrq and PTF\,12dam UVIS observations), we adopted a pixel fraction of unity in the drizzling, while for the others we used a pixel fraction 0.6.

The host-galaxy magnitudes were determined using circular aperture photometry, using zeropoints from the ACS and WFC3 handbook and applying an aperture correction to an infinite aperture.

Some of our {\it HST} observations were conducted within 1--2\,yr after the SN in order to follow the late-time evolution of the light curve, and the SN is still clearly detected.  For the PTF\,10nmn observation, the SN is well offset from the bulk of the host-galaxy light; we subtracted the SN contribution by PSF fitting of the SN using the profile of a nearby star on the same final image with a custom IDL routine.  For PTF\,11hrq, the SN location is also offset from the host galaxy, but it is not clearly detected in the {\it HST} image and no correction is applied. For PTF\,12dam, it is difficult to directly estimate the SN brightness as its location is consistent with a compact but resolved knot of the host galaxy; however, its contribution (relative to the host) is negligible in roughly coeval ground-based photometric and spectroscopic observations.

\begin{figure*}
\centerline{
\includegraphics[scale=0.88,angle=0]{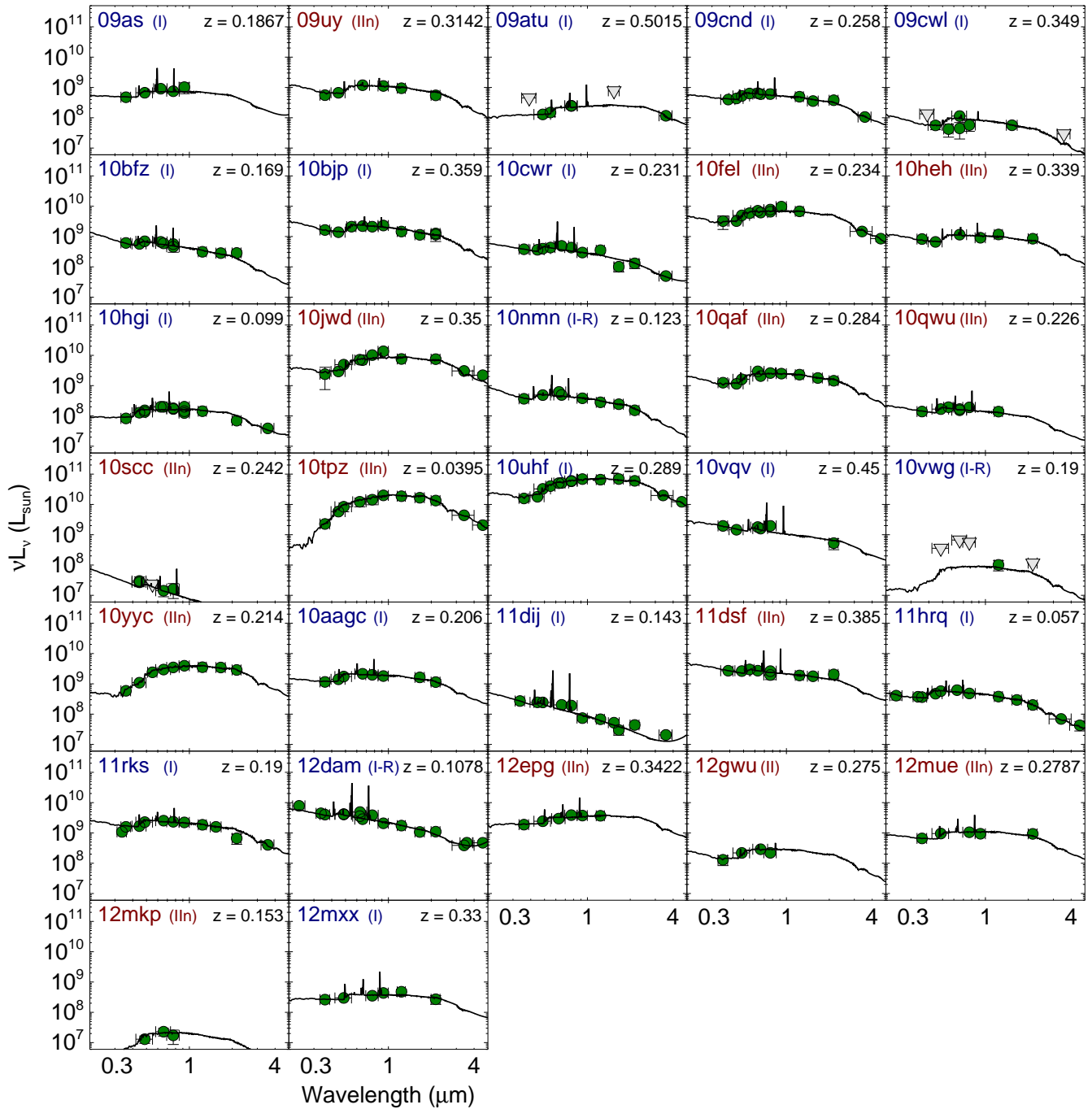}} 
\caption{Spectral luminosity distributions for the host galaxies of all SLSNe in PTF, showing our multiband photometry (green points) and best-fit SED model using the procedure outlined in \ref{sec:sedfit}.  The scale of every subplot is identical, although note that the abscissa is in the observer frame and the ordinate shows the $\nu L_\nu$ luminosity.   Grey triangles indicate upper limits.}
\label{fig:seds}
\end{figure*}

Additional UV and NIR observations of various events in the sample (PTF\,09atu, PTF\,09cnd, PTF\,09cwl, PTF\,11dij, PTF\,11dsf, PTF\,11rks) were obtained as part of programs GO-13025 and GO-13480 (PI A. Levan). Photometry is taken directly from the recent publication of \cite{Angus+2016} with the exception of the F160W photometry of PTF\,11rks, which we recalculated using a larger aperture more appropriate given the diffuse extensions apparent in the ground-based optical imaging.  All {\it HST} photometry is presented alongside our ground-based measurements in Table \ref{tab:photometry} and Figure \ref{fig:seds}.

\subsection{Spitzer and WISE Observations}

Several host galaxies in our sample were observed with IRAC on the {\it Spitzer Space Telescope} during Cycle 10 (GO-10056, PI R. Lunnan).  Sources were observed in channel 1 ($\lambda = 3.6\,\mu$m) using a frame time of 100~s per exposure.  A total integration time of 3600~s was used for each of PTF\,09cnd, PTF\,09atu, and PTF\,09cwl; 1800~s was used for PTF\,10cwr, PTF\,10hgi, and PTF\,11dij, and 1200~s was used for PTF\,11rks and PTF\,12dam.   We downloaded the processed (PBCD) images from the {\it Spitzer} Heritage Archive, subtracted all nearby contaminating sources, and performed aperture photometry on the host galaxy using the procedure outlined by \cite{Perley+2016b}.  Photometry is presented in Table \ref{tab:photometry}.

Many of our galaxies are sufficiently bright to be detected in archival data from the {\it Wide-Field Infrared Survey Explorer} (\textit{WISE}; \citealt{Wright+2010}).  We downloaded photometry from the online catalogs of the ALLWISE Data Release \citep{Cutri+2013}\footnote{\url{http://wise2.ipac.caltech.edu/docs/release/allwise/}} and include them in our tables and SED fits.

\subsection{Spectroscopy}

Spectra were obtained for most host galaxies within the sample using LRIS on the Keck I telescope. We used the 400/8000 grating on the red side and either the 400/3400 or 600/4000 grism on the blue side, providing continuous coverage between the near-UV and 10300\,\AA. 
In cases where our ground-based imaging was able to clearly resolve the host, the slit was aligned with the major axis of the galaxy to minimize slit losses (in all cases a 1$\arcsec$ slit was used); otherwise, the parallactic angle \citep{Filippenko1982} was used.  LRIS is equipped with an atmospheric dispersion corrector \citep{Phillips+2006}, so flux losses associated with nonparallactic angles are minimal.  A few systems (PTF\,10uhf, PTF\,10tpz, PTF\,11rks) exhibit large, resolved hosts which cannot be easily accommodated in the slit; both of these systems appear to be special cases and are discussed individually later.  A log of all exposures is presented in Table \ref{tab:specobslog}.

All LRIS spectra were reduced in a standard manner using the tools in our custom LRIS pipeline LPipe\footnote{\url{http://www.astro.caltech.edu/~dperley/programs/lpipe.html}}, with the extraction aperture carefully determined to include all nebular flux from the host galaxy evident in the two-dimensional frames and exclude flux from neighboring companion objects.

Two host galaxies were observed with DEIMOS \citep{Faber+2003} on the Keck II telescope at the parallactic angle: PTF\,10hgi and PTF\,12epg. The data were reduced within IRAF using a similar general procedure as the LRIS data.

Flux calibration was performed relative to standard stars using the same setup and (where possible) similar airmass.  The portions of the LRIS spectra separated by the blue and red cameras were joined with the flux rescaled based on the measured source plus sky flux within the aperture. We note that this process (based on a relatively small overlap region over which the transmission on both sides varies rapidly with wavelength) produces some additional systematic uncertainties in comparing relative fluxes in the blue versus the red halves of the spectra, particularly for sources observed with the 600/4000 grism (which affords limited wavelength overlap).  For sources with measurable continuum, absolute flux calibration (including correction for slit losses) is performed by calculating synthetic photometry on the reduced spectrum and scaling relative to our photometric measurements.  For many of our targets there is no measurable host continuum or the uncertainty in the continuum flux level is dominated by the sky-continuum background subtraction, so the spectroscopic flux calibration is used directly with no rescaling.  

The DEIMOS spectrum of PTF\,10hgi was obtained only a year after the SN and contains significant SN light; in this case, we calibrate the fluxes by scaling the spectrum to match the H$\alpha$ flux of this object as presented by \cite{Leloudas+2015}.  We also take the [O~II] flux from \cite{Leloudas+2015}, since this line is not covered by the wavelength range of this DEIMOS spectrum.  All other spectra were obtained sufficiently late that the SN was not a significant source of emission.

\begin{figure*}
\centerline{
\includegraphics[width=16.5cm,angle=0]{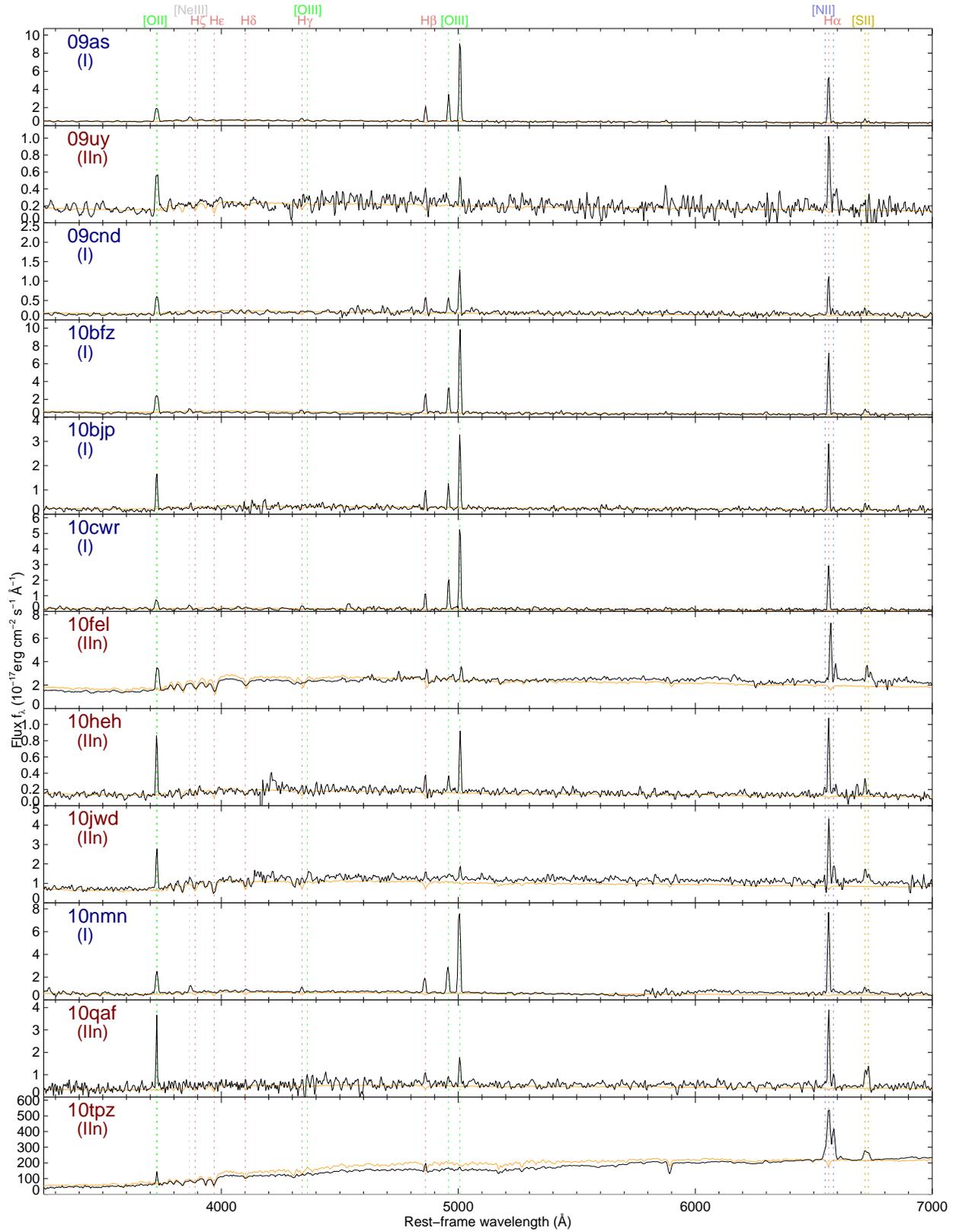}}   
\caption{Keck spectra of the host galaxies of PTF SLSNe with clear continuum and/or line detections. (The spectrum of PTF\,10tpz shows the nuclear region only.)  Overplotted in light orange is the best-fit SED model for the underlying stellar continuum (\S \ref{sec:sedfit}); this may differ somewhat from the observed spectrum for large galaxies where slit-loss effects are significant.}
\label{fig:spectra1}
\end{figure*}

\begin{figure*}
\centerline{
\includegraphics[width=16.5cm,angle=0]{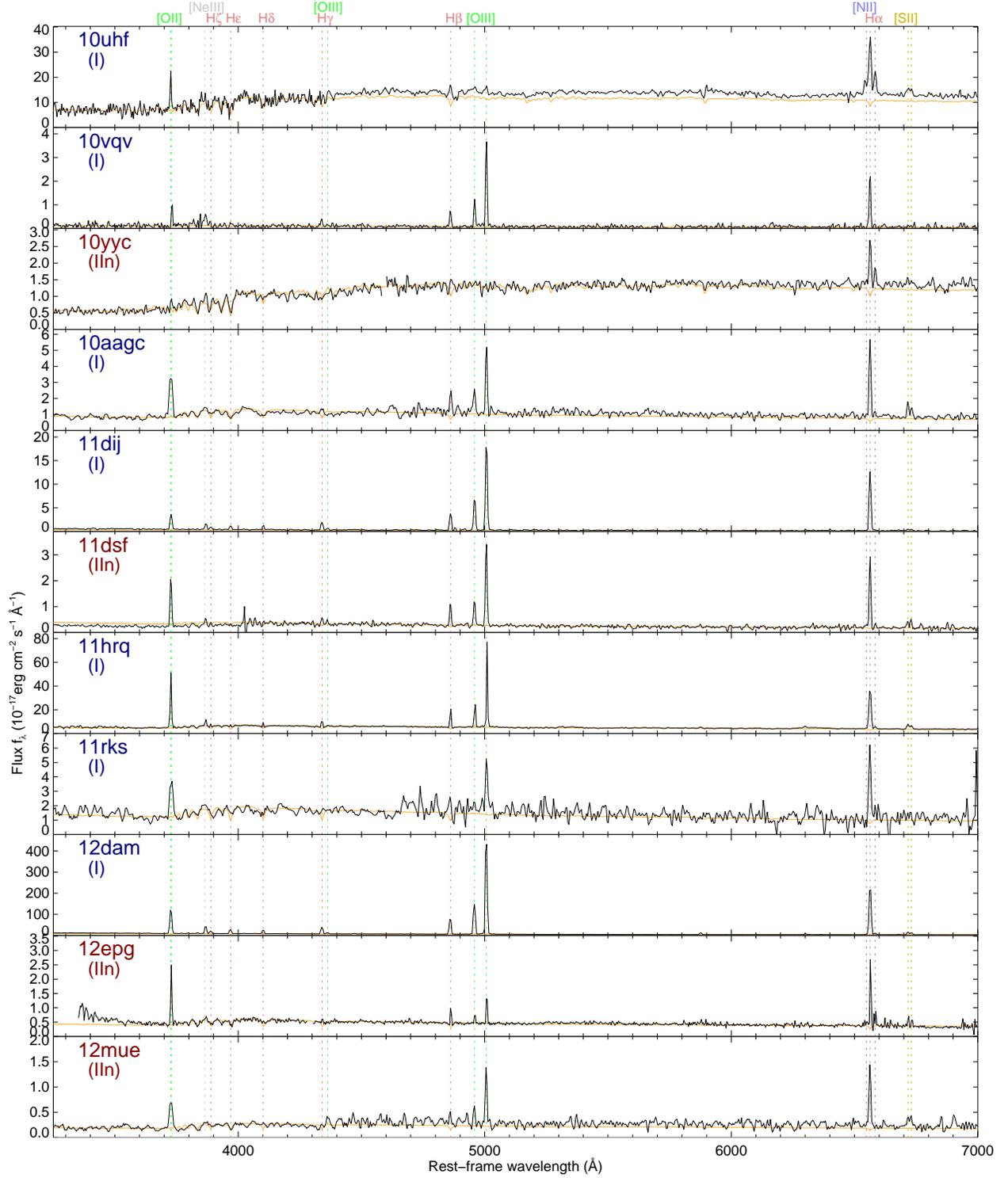}} 
\caption{Additional Keck spectra of the hosts of PTF SLSNe with clear continuum and/or line detections.  
The hosts of PTF\,10uhf and PTF\,11rks are large and resolved; for PTF\,10uhf we show the sum of spectra along two different slit angles (one through each nucleus), whereas for PTF\,11rks we display the spectrum obtained from a slit passing through the southern part of the galaxy (not directly intersecting the galaxy nucleus).}
\label{fig:spectra2}
\end{figure*}

\begin{figure*}
\centerline{
\includegraphics[width=17cm,angle=0]{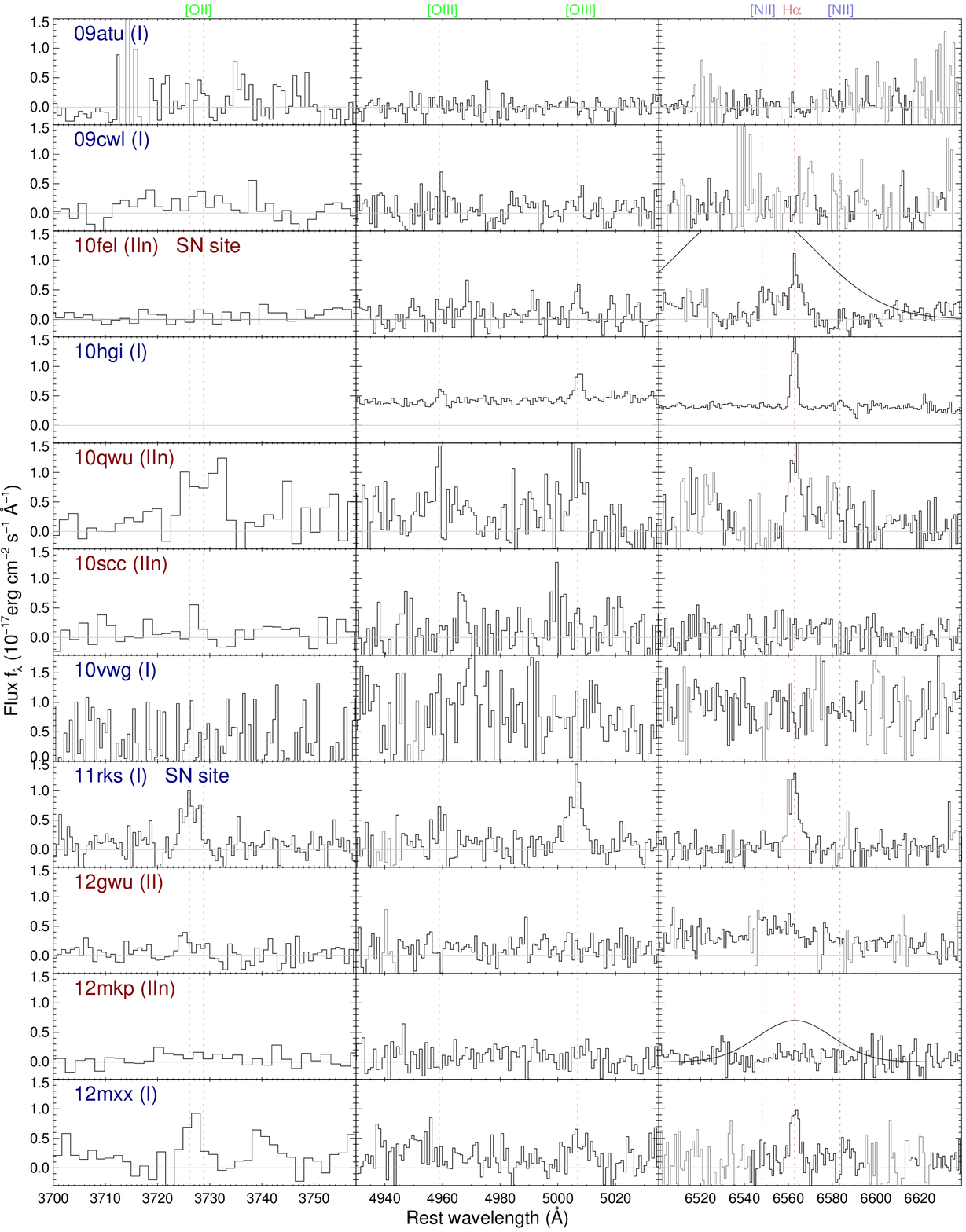}} 
\caption{Spectra of SLSN hosts (or SLSN sites) showing no or very faint line emission at the expected locations of the strongest three emission features in typical host-galaxy spectra: the [O~II] $\lambda{3727}$ doublet (left panel), the [O~III] $\lambda\lambda$4959, 5007 doublet (middle panel), and H$\alpha$ (right panel; locations of [N~II] are also shown).  The Gaussian profiles indicate where late-time broad H$\alpha$ was subtracted from two Type II SLSNe.  }
\label{fig:spectrafaint}
\end{figure*}

Our spectra are plotted in Figures \ref{fig:spectra1}, \ref{fig:spectra2}, and \ref{fig:spectrafaint}, and have been uploaded to WiseREP\footnote{\url{http://wiserep.weizmann.ac.il/}} \citep{Yaron+2012}.

\subsection{SED Fitting}
\label{sec:sedfit}

To estimate the stellar parameters of the host galaxies ($M_*$, SFR, and $A_V$), we analyze the UV-optical-NIR SEDs using a custom SED-fitting code (previously developed by D.~Perley and used in, e.g., \citealt{Perley+2013}).   The code uses the population-synthesis templates of \cite{bc03} summed according to a parametric star-formation history. We assume a \cite{Chabrier2003} initial mass function (IMF) and a stellar metallicity set between 0.2--1.0 Solar---using our best estimate from the spectroscopic analysis where possible, or otherwise estimated from the mass-metallicity relation (see \S \ref{sec:metallicity}) after an initial fit to measure the stellar mass $M_*$.  The contribution of nebular lines to the photometry is included, with the line parameters also fixed using the spectroscopic analysis (see next section).   The star-formation history of each galaxy is fit as a two-population model, with the maximum age of the older population fixed to the age of the Universe at the redshift of the host, and the maximum age of the younger population free (but required to be at least 10\,Myr).  Both populations are assumed to have a continuous star-formation history from the maximum age until today, so the overall star-formation history is constant with an abrupt increase at an arbitrary time $t_{\rm burst}$, the age of the current ongoing starburst.  Dust attenuation is assumed to follow the \cite{Calzetti+2000} template.

For a few host galaxies, we do not have sufficient data to constrain all parameters in the model.  In these cases we fix the burst age to 100\,Myr (this is approximately the timescale to which near-UV luminosities are sensitive, and it enables a reasonable estimate of the average SFR) and/or the extinction $A_V$ to zero, to enable us to still fit the data and obtain reasonable estimates of the SFR and $M_*$.  We also fix the burst age to 100\,Myr if the SED fit converged to a ``negative'' burst corresponding to a decrease in SFR in the recent past.  We apply a continuous star-formation history in cases where the fit permitting an impulsive change converges to a result indistinguishable from this model.

Our results are presented in Figure \ref{fig:seds} and Table \ref{tab:hostproperties}.

\subsection{Spectroscopic Analysis}
\label{sec:specanalysis}

To measure emission-line fluxes from our one-dimensional spectra, we first subtract the host continuum from the SED model (\S \ref{sec:sedfit}) convolved to the resolution of the spectra to remove stellar Balmer absorption, and then fit a Gaussian to the profile of all lines of interest (H$\alpha$, H$\beta$, H$\gamma$, [O~II] $\lambda$3727, [O~III] $\lambda\lambda$4363, 4959, 5007, [N~II]$\lambda\lambda$6548, 6583, [S~II] $\lambda\lambda$6716, 6731).  For faint lines, the width and velocity center of the line model are tied to a nearby strong line: e.g., [N~II] is tied to H$\alpha$.  The components of the [O~II] doublet are always blended at the resolution of our spectra and are reported as a total flux.  These fluxes are given in Table \ref{tab:linefluxes}.  Equivalent widths are calculated at the same time using the measured (pre-subtraction) continuum.

A variety of standard techniques is then used to measure several key physical parameters associated with the fluxes and ratios of nebular lines for each host.  All lines are first corrected for foreground extinction \citep{Schlafly+2011} and converted to luminosities in the host frame.   We directly calculate Balmer decrement extinction, [N~II]/H$\alpha$, $R_{23}$ \citep{Pagel+1979}, and the gas-phase oxygen abundance values (``metallicities'') associated with the latter two using the empirical derivation of \cite{Nagao+2006}.   Electron-temperature ($T_e$) metallicities are also calculated for the subset of hosts in which the auroral [O~III] $\lambda$4363 line is detected using the iterative method of \cite{Izotov+2006}.  Uncertainties in all of these values are calculated by performing 1000 Monte-Carlo trials using the measured flux uncertainties.  In addition, we use the code of \cite{Bianco+2016} to calculate metallicities and Monte-Carlo uncertainties using a range of additional diagnostics and calibrations (specifically, those of \citealt{Denicolo+2002,Pettini+2004,Maiolino+2008,Marino+2013,KK04}; and \citealt{Kewley+2002}).  SFRs are calculated with the standard conversion of \cite{Kennicutt+1987}, adjusted for a Chabrier IMF.

\subsection{Comparison Samples}
\label{sec:catalogs}

In order to help discern to what extent the observed host properties of the SLSN sample reflect the physical influences affecting SLSN production (rather than simply where stars are forming in the low-redshift universe) it is necessary to compare our host galaxies to a representative sample of star-forming galaxies at similar redshifts.  The ideal sample for this purpose would provide all of the same measurements available for our hosts (i.e., integrated UV-through-NIR SEDs and optical spectroscopy, and physical parameters derived from these measurements) and be complete down to a comparable luminosity limit as our sample ($M_g \lesssim -14$ mag, with no targeting or other systematic biases) over a sufficiently large volume to avoid systematics associated with cosmic variance.  Ideally it would also have a similar median redshift ($z\approx0.3$) as our host sample.  

Unfortunately, no single publicly available galaxy sample currently exists satisfying all these criteria simultaneously: most surveys (especially spectroscopic surveys) are complete only to high-mass galaxies and insensitive to the low-mass dwarf galaxies which dominate the SLSN host population.  Certain surveys come close in restricted regimes, or can be made to be (nearly) volume-limited by employing a stringent distance cut or cuts.  We rely on three different surveys in our comparisons, outlined below.

\subsubsection{LVL Galaxies}

The Local Volume Legacy Survey (LVL; \citealt{Dale+2009}; see also \citealt{Kennicutt+2008,Lee+2011,Cook+2014,Cook+2014a}) is a volume-complete sample of galaxies within 11 Mpc.  The catalogs include narrowband-imaging derived $H\alpha$ fluxes and thorough UV-through-NIR photometry.  This catalog has the advantage of being essentially complete down to extremely faint levels for star-forming galaxies, and has very thorough and reliable multiwavelength measurements.  However, the survey volume is small---subjecting it to both statistical limitations (the survey includes only 313 galaxies in total, and rare classes of galaxies are poorly sampled) and systematic/cosmic-variance effects associated with only sampling a region around the Local Group (devoid of large voids or clusters).  It also samples only the $z\approx0$ universe and is purely a photometric survey with no spectroscopy, although metallicity measurements are available for about 2/3 of these galaxies from the literature (we employ the compilations in \citealt{Marble+2010} and \citealt{Cook+2014}, and add measurements for NGC\,1569 and II\,Zw\,40 from \citealt{Kobulnicky+1997} and \citealt{Walsh+1993}, respectively).  Even with these limitations, it is by far the most reliable catalog for examining the contribution of faint, low-mass galaxies to the local SFR density.

\subsubsection{Ultra-VISTA Galaxies}

The Ultra-VISTA catalog \citep{Muzzin+2013} is a very deep multiband photometric survey in the COSMOS field targeted mainly at high-redshift galaxies.  The photometric redshifts obtained by the survey allow it to reach low-mass galaxies at redshifts comparable to our SLSN hosts, although it does not provide $H\alpha$ measurements and it remains incomplete at very low masses: 
specifically, below $3\times10^8\,{\rm M}_\odot$ at $z\approx0.2$.  The SED fitting procedure used also appears unable to recognize starburst galaxies since essentially no galaxies with an SFR greater than $>2\,{\rm M}_\odot$\,yr$^{-1}$ are present within the catalogs at low redshifts (with the exception of a small number of objects which appear to be due to problematic fits since the SED SFRs exceed the direct UV SFRs by an unreasonable factor; these are cleaned from the catalog in our plots/analysis).  Nevertheless, the catalog is large (17932 galaxies at $z<0.6$) and the stellar mass measurements should be robust, so it provides an excellent complement to LVL to check for the effects of cosmic variance or redshift evolution in the photometric data.

\subsubsection{SDSS Galaxies}
\label{sec:sdss}

The Sloan Digital Sky Survey (SDSS) spectroscopic galaxy sample has been extensively used for comparisons to SN and GRB host populations for over a decade, thanks to its extremely wide coverage and high-quality spectroscopy including flux measurements of all the key strong emission lines used for metallicity and other gas-phase analysis (see, e.g., \citealt{Modjaz+2008,Graham+2013}).  Despite these advantages SDSS has several important limitations.  First of all, it is relatively shallow: spectroscopic coverage is complete only to $r < 17.77$ mag \citep{Strauss+2002}, so it is insensitive to low-mass galaxies except at very low redshifts.\footnote{This completeness is also limited to objects morphologically classified as galaxies, meaning it may accidentally exclude compact, high-surface brightness galaxies as stars.} 
Because of the survey's vast areal coverage, it can simply be restricted to quite low redshifts (where it is complete to lower-mass galaxies) and still maintain a quite large sample size.   However, at low redshifts two other challenges come into effect.  First, the wavelength range of the survey does not cover the [O~II] line (needed for $R_{23}$-based metallicity derivations or ionization-parameter analysis) at redshifts below $z<0.021$.  Even more importantly, SDSS spectroscopy is based on fibers of limited area: 3$\arcsec$ diameter or 1.5$\arcsec$ in radius, meaning that at low redshifts essentially only the galaxy nucleus is covered ($r < 300$pc at $z=0.01$).  As a result, a significant tension emerges---the high-$z$ subset is incomplete to the low galaxy masses relevant to our SLSN hosts, but the low-$z$ subset is highly fiber-biased, causing it to (in particular) overestimate metal abundances and underestimate SFRs due to systematic differences between the nucleus and the disk in typical large galaxies.

We attempted to mitigate these issues by adopting a hybrid approach to selection, using different redshift ranges to sample different mass ranges.  As our base catalog, we utilize the NASA-Sloan Atlas (NS-Atlas\footnote{\url{http://www.nsatlas.org/}}) reanalysis of the SDSS spectroscopic sample, a re-reduction of the main galaxy sample (at $z<0.05$) with improved sky subtraction and spectroscopic analysis, as well as mass measurements based on SED fitting to the GALEX+$ugriz$ photometry.  We then take five different mass-redshift slices (each of the same cosmic volume): $z=0.0025-0.0085$ for $M_*<10^7$, $z=0.005-0.009$ for $10^7<M_*<2\times10^8$, $z=0.01-0.0117$ for $2\times10^8<M_*<1\times10^9$, and $z=0.025-0.02532$ for $10^9<M_*<5\times10^9$, and $z=0.0450-0.0451$ for $M_* > 10^9$, leaving a final sample of 1497 galaxies complete down to $\sim10^7\, M_\odot$.  Even with these corrections, the fiber covers only the central region of each host ($r<$50, 100, 200, 500, and 880 pc, respectively, for each redshift bin), and metallicity- and specific SFR-gradient effects are likely to be large.  We therefore generally favor the smaller LVL sample in our analysis, except in cases where knowing the individual measured line fluxes is necessary (\S \ref{sec:lineratios}).

\section{Discussion of Individual Objects}
\label{sec:individuals}

\subsection{PTF\,09as}
This event was found very early in the survey but was initially misclassified as a Type Ia SN and not recognized as a SLSN until 2015 (six years after discovery) after re-analysis of archival PTF spectra (Quimby et al., in prep).  Accordingly we have no late-time imaging, but the host is detected in SDSS pre-imaging (in $g$, $r$, $i$, and marginally in $z$); the host appears only marginally resolved.  Our LRIS spectrum of this galaxy shows a modest continuum and strong emission from all major nebular lines.

\subsection{PTF\,09uy}
A moderately faint, unresolved source is visible underlying the SN position in late-time MOSFIRE and LRIS imaging which we identify as the host galaxy.  Spectroscopy shows emission lines of [O~II], [O~III], H$\alpha$, [N~II], and [S~II] as well as strong continuum from a more mature stellar population.  There is no evidence of an AGN contributing to the flux, supporting our identification of the transient as a SLSN.

\subsection{PTF\,09atu}
Discovery imaging and classification spectra of this event were previously presented in \cite{Quimby+2011}, along with those of PTF\,09cnd, PTF\,09cwl, and PTF\,10cwr.  It is the most distant event in our sample, with a redshift of $z=0.501$ (well-determined by narrow Mg~II absorption in the original SN spectra; \citealt{Quimby+2011}).
 Late-time imaging with Keck and {\it HST} detect a persistent source within 1$\arcsec$ of the SN location, but it is very faint ($R\approx26$ mag) and the detections in each filter are individually marginal. Cumulatively, the detections are significant and we identify this target as the probable host galaxy, although given its offset and faintness we cannot completely rule out that it is a chance coincidence (in which case the host would be even fainter.)
Keck spectroscopy over the putative host location shows neither any detectable continuum nor any lines at the expected wavelengths of strong galaxy emission features given the SN redshift (or elsewhere in the spectrum), indicating a low-mass galaxy not in a starburst phase.

\subsection{PTF\,09cnd}
The SN is located on the northeast fringe of a compact, moderately bright galaxy visible in our late-time multicolor imaging.    Spectroscopy reveals a strong continuum with various strong emission features superimposed at a redshift consistent with that of the narrow Mg~II absorption in the early-time spectrum.   A neighboring source (2.2$\arcsec$ from the host) is also seen in our imaging; the slit was aligned to cover both sources but the companion shows no strong emission or absorption lines, so its redshift or association with the host is unknown.

\subsection{PTF\,09cwl (SN\,2009jh)}
The host galaxy of this event is extremely faint ($R\approx26$ mag).  No significant flux at the wavelengths of any expected strong emission line features is evident in our spectroscopy.

\subsection{PTF\,10bfz}
The host of this event is well-detected in our imaging, is unresolved, and there are no nearby companions.  Strong nebular emission lines are evident, indicating a compact, low-mass galaxy undergoing active star formation.

\subsection{PTF\,10bjp}
The appearance of two distinct sources connected by a bridge of emission strongly suggests that this host galaxy is undergoing a major merger; spectroscopy confirms that they are at the same redshift.  Both are strongly star-forming.  We report the combined fluxes of both galaxies (and treat the two galaxies as a single object) throughout the paper.  The SN occurred in the southern member of the pair.

\subsection{PTF\,10cwr (SN\,2010gx)}
The host galaxy is bright and compact in our imaging and directly underlies the SN position, although a second object is easily visible in the imaging $\sim 1\arcsec$ east of this location.  While our late-time host galaxy spectrum was not oriented to cover this object, an earlier spectrum of the SN did place this object on the slit and shows it to be a star-forming galaxy in the background at $z=0.622$ and therefore unassociated.  The host itself is strongly star-forming and shows a clear detection of the [O~III] $\lambda$4363 auroral line, marking it as quite metal-poor:  indeed, it is the most metal-poor galaxy among all members of our sample for which spectroscopic measurements are available.  (These properties were also remarked upon by the dedicated study of \citealt{Chen+2013}.)

\subsection{PTF\,10fel}
\label{sec:10fel}
Imaging of the field shows a complex region surrounding the transient location.  A massive disk galaxy is evident $\sim 4\arcsec$ southeast in projection from the SN location (with the offset nearly aligned with the galaxy's major axis); two faint point-like sources are visible $\sim 1\arcsec$ north and $\sim 4\arcsec$ northwest of the SN location (respectively), and some faint, diffuse emission surrounds the entire region.  The SN itself is not coincident with any of the individual objects.   

A late-time spectrum aligned with the SN location and the closer of the two point sources shows no detectable emission at the SN site but weak emission from the point source at a much higher redshift, indicating that this source is unassociated.  An earlier, deeper spectrum of the SN itself shows faint narrow H$\alpha$ emission at the SN location after the broad H$\alpha$ line from the SN is subtracted.

We treat the nearby disk galaxy as the ``host'' for our purposes---although it is possible that the SN is actually associated with an extremely faint unseen satellite.  Even if the SN is associated with this galaxy, it evidently occurred in a location that is not representative of the star-forming conditions probed by our spectroscopy of that galaxy, and so the properties we report for this system should be used with caution when drawing conclusions about the host population.

\subsection{PTF\,10heh}

The host is not significantly resolved by our imaging observations.  It shows a weak continuum and strong emission features, indicating an actively star-forming low-mass galaxy.

\subsection{PTF\,10hgi}

This is one of the most nearby events in the sample at $z=0.098$.  The host is well-detected in ground-based imaging and resolved into a disk-like structure, although it has a very low luminosity ($M_g \approx -16$ mag). The emission lines from this galaxy are quite weak and it clearly demonstrates that SLSNe can form in dwarf galaxies outside of their starburst phases.  It may be a good analog of more distant (and harder-to-study) low-luminosity hosts such as PTF\,09atu and PTF\,09cwl.

\subsection{PTF\,10jwd}
The host of this event is quite bright; morphologically this event resembles a resolved, nearly edge-on disk similar to PTF\,10hgi with a hint of some asymmetry.    The spectrum is dominated by an older population with a clear Balmer break, although significant emission lines are present also---including moderately strong [N~II] and [S~II].

\subsection{PTF\,10nmn}

Because the redshift of this event places several of the critical strong nebular emission lines near the half-power point of the LRIS D560 dichroic, this host galaxy was instead observed using the lower-resolution B300 grating and the D680 dichroic.  Unfortunately this introduces some additional complications associated with second-order light (both the host and the standard star observed in this setup are relatively blue), complicating the flux-calibration.  This mainly affects the region of the spectrum between approximately 6000-6600 \AA\ (5300--5900 in the host frame) where there are no strong lines of interest, although it does impede accurate calculation of the relative scaling between lines redder than this (H$\alpha$, [N~II], and [S~II]) and the remaining lines.

The host is small and only marginally resolved by ground-based imaging, which shows a central core with extensions in various directions.  The {\it HST} imaging resolves the core into a diffuse structure and situates the SN about one arcsecond southwest of the galaxy center, along a narrow extended structure (possibly a tidal feature).  The LRIS slit was aligned with the host center and the SN site; while strong nebular emission is present throughout the host along this slit, all emission lines are much stronger in and around the location of the SN site while the underlying continuum is much weaker, indicating that the SN occurred in a particularly young and active star-forming region.

To be consistent with the rest of our analysis, and because the star-forming region at the SN site dominates the nebular emission of the host overall, we extract the spectrum of the entire host galaxy.  The marginally resolved nature of the host permits some spatially resolved analysis that will be reported by Yaron et al.\ in forthcoming work.

\subsection{PTF\,10qaf}

PTF\,10qaf occurred in a dwarf companion to a large, face-on spiral galaxy with which it is presumably interacting (our spectroscopy confirms it to be at the same redshift), although the two galaxies are physically separate and do not appear to be actively merging.
The host has previously been studied by \cite{Leloudas+2015}, although they do not distinguish that the host and its neighbor are separate galaxies (they do extract the host separately in their spectroscopy and refer to it as the SN site).   The host has strong emission lines, but is not remarkable and the strong detections of [N~II] and [S~II] (and relatively weak [O~III] $\lambda\lambda$4959, 5007) suggest that it is neither starbursting nor particularly metal-poor.

\subsection{PTF\,10qwu}
The field surrounding the location of 10qwu is complex.  Three unresolved/marginally resolved sources appear close to the SN site: the faintest of these is coincident with the SN, a slightly brighter source is separated by 1 arcsec to the southeast, and a second, even brighter source is separated by about 2 arcsec to the southwest.  The slit was oriented to cover the fainter two sources.  We detect no continuum or line emission at the faintest of these (the SN site).  At the neighboring, brighter source we detect marginal continuum emission and weak lines corresponding to H$\alpha$, [O~II], and [O~III], at a redshift equivalent to that determined from the narrow-line emission from the SN at earlier times.  We treat this object as the host galaxy (with the SN site on a faint extension) although it is possible the two are not connected and only the underlying source is the host galaxy, in which case it would be much fainter than as presented here.

\subsection{PTF\,10scc}

The host galaxy of this event is (along with PTF\,12mkp) the least luminous in the sample: despite a relatively nearby redshift of $z=0.242$ the host is only weakly detected in deep Keck imaging; the inferred absolute magnitude is $M_g =  -13.5$.  As with other very faint hosts, spectroscopy at this location shows no significant emission lines at this position. 

\subsection{PTF\,10tpz}
\label{sec:10tpz}

This SN is unusual in many regards (see also \S \ref{sec:sampleproperties}).  It is the nearest event in the sample by a large margin, it is highly extinguished (without an extinction correction it would seem far from superluminous), and it is located very close to the nucleus of a massive early-type galaxy (adaptive optics imaging, to be presented in future work, demonstrates that it does not originate from the nucleus itself).  These properties call to mind SN\,2006gy (e.g., \citealt{Smith+2007}) and candidate SN CSS100217 \citep{Drake+2011} and are completely distinct from any other host in our sample, possibly suggesting that this event belongs to a distinct progenitor class from the others presented here.  

Both the continuum and nebular flux of the galaxy are dominated by the inner region where the SN occurred, and for this particular host galaxy we extract only this innermost region.\footnote{Because of the very extended nature of this galaxy, most of the light from the disk is not contained within the slit, and extracting the entire galaxy would make only a modest difference to the resulting spectrum but would greatly degrade the S/N.  Accurate characterization of the entire host will require integral-field-unit spectoscopy or other special techniques.}  The [N~II]/H$\alpha$ ratio in this spectrum suggests that an AGN is present, making this the only host galaxy in our sample containing an (observable) AGN.   The very weak [O~III] emission suggests that the AGN is subdominant, however, and the observed nuclear flux likely originates from a mixture of star-formation and AGN activity \citep{Bamford+2008}.  We ignore this unknown AGN contribution in our line analysis, but emphasize that [N~II]-based estimates of the metallicity are unreliable and the true H$\alpha$ SFR will be somewhat less than our analysis implies.

\subsection{PTF\,10uhf}

At a glance this host appears unique in our sample (see Figure 1) as the only large grand-design spiral.  The SN site is located 4$\arcsec$ (17.5 kpc) from the host's center in an extended arcing feature resembling a spiral arm.  However, careful examination of the images reveals a second nucleus 2 arcsec northwest of the primary nucleus and a second pair of ``arms'' with a similar sigmoid shape as those of the primary spiral.  Most likely this sytem represents the merger of a massive spiral galaxy with a somewhat less massive disk galaxy, with both objects pulled into a distorted shape by the interaction.  The SN itself cannot be unambiguously associated with either system; its location is at the overlap of the northern ``arms'' of both galaxies.

We obtained spectroscopy with the slit oriented between the SN site and the nucleus of the ``major'' galaxy, as well as with the slit aligned to cover the ``minor'' merging companion in the interaction.  Nebular emission is detected throughout the positions of the galaxy on both slits including at the SN site.  Due to the difficulty of concretely associating the SN with either component individualy and the fact that the galaxy pair would be treated as single by most surveys, we treat the two galaxies as a single object and report their properties together (using a large aperture, and summing our slits along both axes).  We also extracted a spectrum at the SN location exclusively, excluding the galaxy nuclei.

While the star-formation intensity is moderate throughout (e.g., the [O~III] line is weak throughout both slits including at the SN site) the cumulative flux of H$\alpha$ and [O~II] are very high and the H$\alpha$/H$\beta$ ratio indicates a significant extinction correction, making this host both the most massive and the most rapidly star-forming in the sample by a large margin (it is probably a luminous infrared galaxy [LIRG]).  Indeed, it appears as an outlier in almost every parameter when compared to other SLSN-I hosts---which may suggest that it may be more reasonable to attempt to associate the SN with the ``minor'' component exclusively.  However, even if we made this association, it would reduce the mass and SFR only by approximately a factor of 3--4 (and not significantly alter the metallicity or other line parameters) and it would remain an outlier.  

\subsection{PTF\,10vqv}
The host of PTF\,10vqv is isolated and generally unresolved, although a faint extension is visible to the south.  The SN site is coincident with the host's brightest region.  The spectrum shows weak continuum and very strong star-formation dominated by intense [O~III] emission.

\subsection{PTF\,10vwg (SN\,2010hy)}
This event was discovered by KAIT/LOSS (and observed by PTF) in a moderately crowded low-Galactic latitude field with significant foreground extinction ($A_V = 1.45$\,mag).
Our initial ground-based optical imaging (in relatively poor seeing conditions of 1.1$\arcsec$) suggested that the event originated in the apparent outer regions of a diffuse structure, which appeared at the time to be a candidate host galaxy.  Subsequent imaging with MOSFIRE under superior seeing conditions and image quality resolved this source into three separate point sources---most likely, foreground Galactic stars---all of which are separated from the host (the offset of the nearest source is approximately 0.8$\arcsec$).  Faint emission can be seen under the SN location only in $J$-band, which presumably represents a very faint ``true'' host galaxy.  A spectrum with the slit oriented across the optically blended stars and the SN location shows no detectable emission features at any point.  Most likely this event represents an underluminous, non-intensely-star-forming host similar to several of the others discussed above, although the optical blending (and foreground extinction) prevents equivalently deep limits to be provided.

\subsection{PTF\,10yyc}
Imaging of this field shows two bright, resolved disk galaxies with a similar orientation separated by about 2$\arcsec$.  The SN site is coincident with the outer part of the western, fainter (and redder) galaxy.  We obtained a spectrum oriented across the nuclei of both galaxies.  Despite the small offset they are not at the same redshift and are unassociated: the host of the SN is at $z=0.2146$ (consistent with the SN redshift) while the eastern source is in the foreground at $z=0.198$.  The host has a relatively old stellar population and is only modestly star-forming, with strong Balmer absorption lines, weak H$\alpha$, [O~II], [N~II], and [S~II], and only marginal [O~III] emission.

\subsection{PTF\,10aagc}
The host of this SN is quite bright and shows clear morphological structure at the resolution of our ground-based imaging---suggesting a possible ongoing merger, although the emission lines are only weakly resolved ($<$100 km s$^{-1}$) and the morphology could also be interpreted as a particularly asymmetric dwarf spiral galaxy.  Strong emission lines and continuum are both present in the spectra.

\subsection{PTF\,11dij (SN\,2011ke)}
This source is (along with PTF\,12dam) a classic example of the ``extreme emission-line galaxies'' discussed among the SLSN-I hosts in \citealt{Leloudas+2015} (and \citealt{Lunnan+2014}); it has already been discussed in some detail in these works as an intensely star-forming, low-metallicity, compact galaxy.   It is part of a small group, with a disk galaxy offset by a few arcseconds\footnote{This disk galaxy is referred to as a ``tadpole tail'' and treated as part of the host system by \cite{Leloudas+2015}, but it appears to be a separate object.  Spiral arms are visible in our $g$-band image, suggesting it is a highly inclined spiral.  The true host is likely a satellite of this object; spectroscopy confirms the two galaxies to be at consistent redshifts.}, although there are no obvious signatures of interaction in the morphology.

\subsection{PTF\,11dsf}
The host is compact and isolated, and relatively bright.  The 2D spectrum shows numerous strong nebular emission lines; their relative fluxes are fully consistent with a star-forming galaxy with no significant AGN contribution.  {\it HST} imaging resolves the host into a small irregular galaxy with no clear point-source nucleus \citep{Angus+2016} and the transient is located off-center, further supporting the SN interpretation of the transient.

\subsection{PTF\,11hrq}
This is the nearest event in the sample (excluding the circumnuclear PTF\,10tpz).  The host is well-resolved in the imaging and is unremarkable, appearing as a partially edge-on disk with no obvious structure.  The host exhibits strong emission lines, but is not particularly young or starbursting: high-order Balmer absorption features are evident in the continuum and the line equivalent widths are moderate. 

\subsection{PTF\,11rks}
The large galaxy nearest PTF\,11rks (at the same redshift, and identified as the host by, e.g., \citealt{Lunnan+2014} and \citealt{Angus+2016}) is an extended object with hints of complex structure that may suggest a post-merger system or an otherwise turbulent recent history, although it is not particularly actively star-forming anywhere.  The SN site is highly offset (by 2.8$\arcsec$, or 9.2 kpc) from the nucleus, located on top of an extended ($\sim4.5\arcsec$ or 15 kpc long) linear feature in the ground-based imaging.  The nature of this feature is unclear: it could be an outer spiral arm, or a lower-mass companion galaxy seen edge-on.  A zeroth order image associated with an {\it HST} grism observation (Quimby et al., in prep) shows a faint arc connecting this feature to the host nucleus reminiscent of a spiral arm, so it is reasonable to assume that it is ``part'' of the nearby galaxy.  We will follow earlier authors in treating that galaxy as the host, but it is also quite possible that these are tidal tails produced by interaction between two physically distinct galaxies.  The large (``host'') galaxy is moderate-mass, weakly star-forming, and not metal-poor.

We oriented our slit through the SN site, along the direction of the linear feature.  We detect faint stellar continuum, along with weak emission lines of H$\alpha$, [O~III], and [O~II] that are seen only at the SN site and not elsewhere in the companion.   The [O~III]/[O~II] ratio is modest and does not suggest a particularly intense starburst.   The redshift is the same as that of the nearby large galaxy.   Since the slit does not cover any other part of the ``host'' in this orientation, we use an earlier spectrum of the SN whose position angle intersected the galaxy just south of its nucleus for our more detailed analysis.

\subsection{PTF\,12dam}
PTF\,12dam is among the closest SLSNe to date and took place in a particularly intensely star-forming galaxy with very strong emission lines, and as a result this event has already been the subject of a variety of studies \citep{Lunnan+2014,Chen+2015,Leloudas+2015,Thoene+2015}. The stellar population is quite young and the host metal-poor.

To supplement the existing data we provide {\it HST} UV imaging as well as deep ground-based optical imaging.  The inner portion of the galaxy resolves into a series of star-forming knots, while the deep optical imaging shows low surface-brightness extensions in various directions, most prominently a large diffuse component to the southeast but fainter arcs to the north and west---suggestive of tidal features in the advanced phase of a merger.

\subsection{PTF\,12epg}
The host appears compact in our imaging with no close companions.  The host is rapidly star-forming and its spectrum shows bright underlying continuum and nebular lines, although its properties are not extreme.

\subsection{PTF\,12gwu}
The apparent host galaxy is compact and isolated in our imaging.    No fully SN-uncontaminated spectrum is available, but a deep observation late in the nebular phase detects weak continuum with a very faint, broad H$\alpha$ SN feature superimposed.  No narrow emission lines from the host are present at the wavelength of H$\alpha$ or anywhere else in the spectrum.

\subsection{PTF\,12mkp}
The host is unresolved and faint in our late-time imaging.  Two much brighter galaxies are visible in the image at an offset of a few arcsec; we aligned our slit with the nearer of the galaxies and find it to be at a different redshift and therefore unassociated.  Broad H$\alpha$ emission from the SN was still present as of January 2015 at the time of our last spectrum (this is subtracted via a Gaussian fit in Figure \ref{fig:spectrafaint} for clarity), but no narrow host feature is superimposed, nor are other strong emission features detected.

\begin{figure*}
\centerline{
\includegraphics[width=14cm,angle=0]{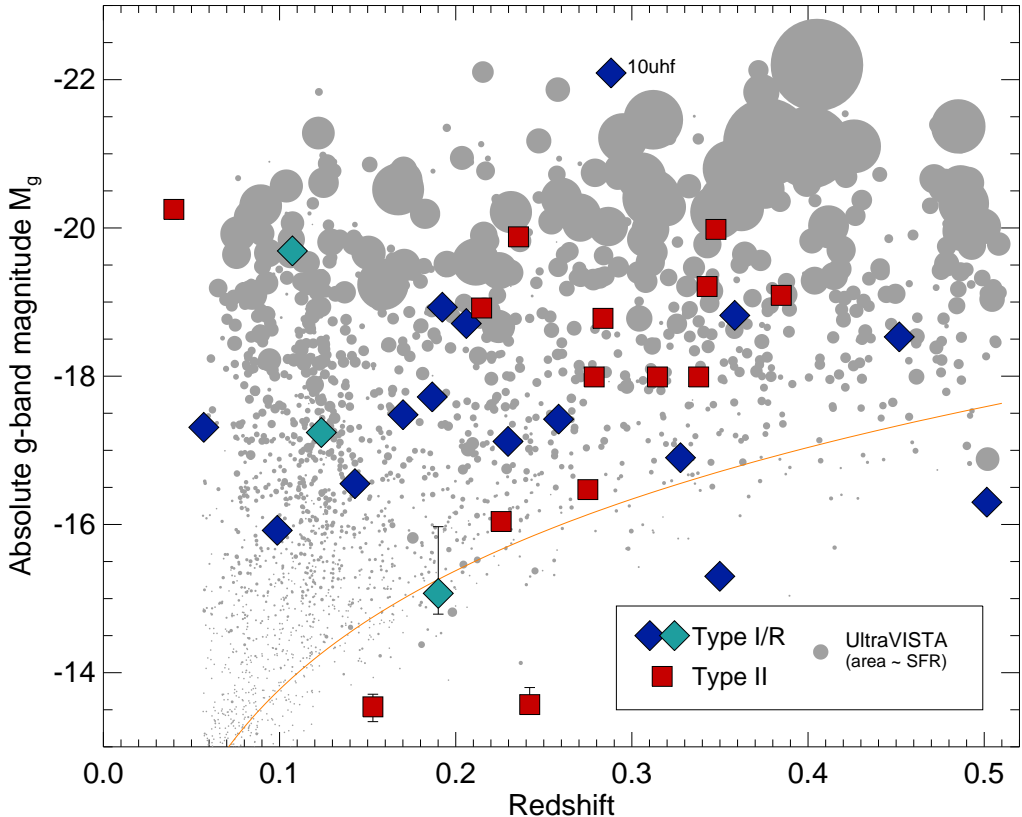}} 
\caption{Absolute $g-$band magnitudes of SLSN host galaxies from PTF as a function of redshift, determined by synthetic photometry of our best-fit model SEDs (\S \ref{sec:sedfit}).  A large sample of star-forming galaxies from Ultra-VISTA is shown in grey, with their sizes scaled according to SFR.  The orange curve indicates an apparent magnitude of $m=24.5$, roughly marking the optical completeness threshold of Ultra-VISTA.   The SLSN-I host population is dominated by low-luminosity galaxies at all redshifts.}
\label{fig:z_luminosity}
\end{figure*}

\subsection{PTF\,12mue}
The host is compact and isolated.  Strong emission lines are present on top of a significant continuum from a more evolved population.

\subsection{PTF\,12mxx}
The host of PTF\,12mxx is faint and not clearly resolved at the resolution of ground-based imaging.  In our spectroscopy we detect weak lines of H$\alpha$, [O~III], [O~II], and marginal continuum.

\vskip 0.7cm

\section{Results}
\label{sec:results}

The host properties of the SLSN sample are broad, spanning the full range experienced by local star-forming galaxies.
Our primary science objective---understanding whether SLSNe preferentially occur in certain types of galaxy over others (and what these preferences are)---requires quantative comparison of the distributions of these physical parameters versus the \emph{expected} distributions as inferred from observations of large, complete samples of star-forming galaxies.  This is easier (and more physically informative) for some parameters compared to others, and we will begin this section by qualitatively examining some basic observational properties before proceeding to more quantitative comparisons of the key physical parameters that might constrain SLSN progenitor models.

\subsection{Detectability, Luminosity, and Color}


The term ``hostless'' is frequently applied to individual SLSNe, indicating that they do not show any evidence of a host galaxy in pre- or post-imaging.  Of course, it would be extremely surprising if any SLSN was truly hostless: their large ejecta masses \citep{Nicholl+2015} require an association with massive stars which cannot feasibly travel significant distances from the environment in which they (and, unavoidably, large numbers of other stars) would have formed.  Furthermore, the detection of weak narrow Mg II absorption (at higher redshifts where this is possible with ground-based spectroscopy; \citealt{Quimby+2011,Vreeswijk+2014}) implies at least some pre-existing circumgalactic matter.

While many of the host galaxies in our sample are exceptionally low in luminosity, our observations confirm that all SLSNe are indeed hosted within galaxies.   Perhaps more usefully, our observations also establish the luminosity scale of several of the faintest SLSN hosts which previously had avoided detection: the faintest hosts in our sample (10scc and 12mkp, both of Type II) have absolute magnitudes of a remarkable $M_g \approx -13.5$ (Figure \ref{fig:z_luminosity}), putting them in the regime of the faintest known actively star-forming galaxies anywhere in the universe.

The association of SLSNe with massive stars also implies that all of their hosts should have active star formation.  The lack of detectable emission lines from many of our systems might therefore be seen as surprising.  However, these nondetections all correspond to the faintest galaxies in the sample, and the limits on the equivalent widths of these lines by comparing our flux limits to the continuum estimated from the photometry does not rule out that they are star-forming at a steady (nonbursting) rate.  Indeed, the broadband colors of all of our ultra-faint hosts are blue, indicating that young ($\lesssim100$\,Myr) stars are present and that steady, low-level star formation is occurring.

Luminous host galaxies are rare in the sample, particularly among the Type I SLSNe.   Only two SLSN-I hosts have luminosities exceeding $M_g \approx -19$ mag: the host of PTF\,12dam (a very young and intense starburst that is not at all representative of luminous galaxies) and the host of PTF\,10uhf (an outlier in our sample and indeed one of the most luminous galaxies in the entire low-redshift universe).  Both these systems will be discussed in more detail later.

Quantitative comparisons against the expected parameter distributions for an SN population that traces star formation uniformly will be deferred to further sections.  We remark, for now, that simple visual inspection of the vertical positions of Type I SLSNe in Figure \ref{fig:z_luminosity} demonstrates that these hosts are not at all typical of the sites of most cosmic star formation (grey points; the area is scaled in proportion to SFR as a visual indicator of their probability of producing a SN per unit time).  Cosmic star formation in the local universe is dominated by high-luminosity ($M \approx -20$ mag) galaxies, confirming that the preference for ``faint'' galaxies (first noted by \citealt{Neill+2011} and also remarked upon by nearly all subsequent work on the topic) is a physical effect evident even in complete single-survey samples.

\subsection{Morphology}

As our observations were conducted (primarily) by ground-based observatories at a variety of wavelengths and seeing conditions (and many of the hosts are not resolved) it is not straightforward to classify the sample morphologically or easily quantify the host sizes; nor is there an appropriate volume-limited comparison sample to which such measurements could be easily compared.  Even so, it is worth briefly and informally remarking on the morphological properties of our sample. 

Most SLSN hosts are small galaxies: inspection of Figure \ref{fig:mosaic} demonstrates that most of our sources are marginally resolved at best, indicating that nearly all of the flux is within a few kpc of the center (typical seeing conditions of 1$\arcsec$ correspond to $\sim2$ kpc at $z=0.1$ or $\sim6$ kpc at $z=0.5$).  This is consistent with the previous results of \cite{Lunnan+2015}, who measured a median half-light radius of only 0.9~kpc for SLSN-I hosts in the Pan-STARRS sample.  Exceptions, where they exist, are fairly dramatic: PTF\,10tpz, PTF\,10uhf, PTF\,10fel, and PTF\,11rks are all very large galaxies spanning tens of kpc in diameter and in all of these cases except the first the SLSN site is highly offset from the nucleus.  

Among the sources that are resolved, mergers are quite common:  the hosts of PTF\,10bjp, PTF\,10cwr, PTF\,10uhf, PTF\,11rks, PTF\,11dij, and PTF\,12dam all show binarity, companions, and/or probable tidal features.  PTF\,10aagc may also be a merger.  All of these are SLSNe-I; no mergers are evident in the SLSN-II sample (note that the apparent companion of PTF\,10yyc is actually a foreground galaxy).  

\subsection{Stellar Masses}
\label{sec:masses}

\begin{figure*}
\centerline{
\includegraphics[width=17cm,angle=0]{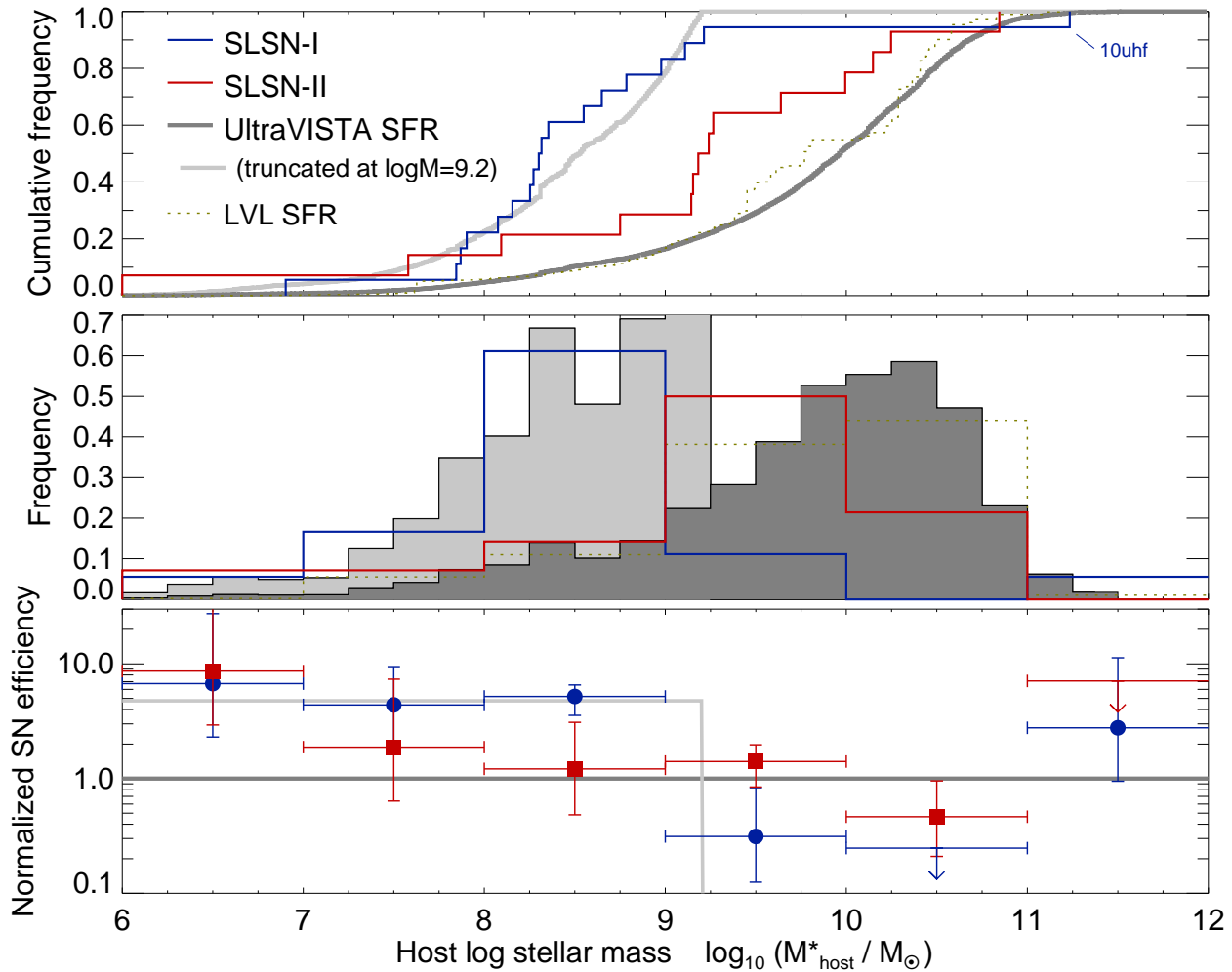}} 
\caption{The SLSN host-galaxy mass distribution of our sample, compared to the cosmic SFR distribution as a function of $M_*$ measured by galaxy surveys (Ultra-VISTA and LVL).  Top panel: cumulative distributions.   Middle panel: histograms, expressed as a frequency per 1\,dex bin in $M_*$.  Bottom panel: Relative SLSN production efficiency per unit SFR, calculated by dividing the SLSN frequency histogram by the SFR frequency histogram and normalizing to an average value of 1.0.   In all three plots, the light-grey curve shows a truncated sample from Ultra-VISTA in which galaxies with log$_{10}(M*/{\rm M}_\odot) > 9.2$ are neglected.   The SLSN-I mass distribution shows good consistency with this model, suggesting that these events are strongly suppressed in high-mass galaxies (but their production efficiency is independent of host mass below $\sim10^{9.2} M_\odot$.)  The efficiency of producing Type II SLSNe does not show strong dependence on stellar mass.
}
\label{fig:massrate}
\end{figure*}

Given their characteristically low luminosities, it is not surprising that SLSN-I hosts are also strikingly low in mass.  The median SLSN-I host mass is only $2 \times 10^8\,{\rm M}_\odot$ (less than the mass of the SMC), and every SLSN-I host in the sample except that of PTF\,10uhf is in a galaxy with $M_*<10^{9.5}\,{\rm M}_\odot$ (approximately the mass of the LMC).  
Type II SLSNe are also found in very low-mass hosts on occasion, but in contrast to SLSNe-I their hosts appear to populate the entire mass distribution of star-forming galaxies fairly uniformly, including the most massive end (although note our remarks about the local environments of PTF\,10tpz and PTF\,10fel; \S\ref{sec:10tpz} and \S\ref{sec:10fel}).

The distribution of cosmic star formation with respect to galaxy stellar mass can be evaluated by binning all galaxies from a volume-limited comparison survey in log-mass space, each weighted by their measured SFRs.  The Ultra-VISTA catalog provides the best comparison sample for this purpose (due to its large volume and the ability to target the same redshift space at which our SLSNe are actually found), although it does suffer the difficulty that at high redshifts it is incomplete to low-mass galaxies (this effect can be easily seen in Figure \ref{fig:z_luminosity}.)  In order to sample the full mass and volume range accessed by Ultra-VISTA we divide the sample into two segments: specifically, we use a redshift range of $z=0.06-0.12$ to sample the lowest-mass ($<6\times10^8\, {\rm M}_\odot$) galaxies (Ultra-VISTA is complete to $7\times10^7\,{\rm M}_\odot$ over this range) and $z=0.12-0.5$ to sample the higher-mass ($>6\times10^8\, {\rm M}_\odot$) galaxies, and reweight the SFRs appropriately to correct for the different volumes sampled.    

The resulting distribution is plotted in the middle panel of Figure \ref{fig:massrate} as the filled histogram.  Low-redshift star formation is dominated by galaxies within about 1 dex of $10^{10}\, {\rm M}_\odot$, with a tail extending to lower masses (in agreement with previous studies, e.g., \citealt{Brinchmann+2004}).  As a check on our procedure we also plot the distribution of the LVL sample and find it to be consistent with the Ultra-VISTA sample (dotted line), demonstrating the effectiveness of our scaling procedure and confirming that the mass threshold of our Ultra-VISTA sample roughly matches the ``real'' mass threshold below which the contribution to current star formation is no longer significant.  Likewise, this demonstrates that redshift evolution over this period is relatively minor.\footnote{Notably, however, the high-$z$ Ultra-VISTA sample contains a small but significant contribution from very high-mass ($>10^{11}\, {\rm M}_\odot$ galaxies); in local samples the contribution from this range is essentially zero.}

This distribution can be compared directly to the mass distributions of our host galaxies, which are also plotted as the solid, colored (and unfilled) curves.  The peak of the SLSN-I host mass distribution at $10^8-10^9\, {\rm M}_\odot$ is clearly much lower than what would be expected for a transient population that traces low-$z$ star formation.  This can be expressed statistically by comparing the cumulative distributions, shown in the top panel: a two-sided K-S test gives a probability of $2\times10^{-9}$ that the SLSNe-I are drawn uniformly from cosmic star formation.  The distribution of SLSN-II hosts is much less concentrated towards low-mass galaxies, but does exhibit a more modest trend towards low masses that is marginally significant (1.0 $\times10^{-2}$ of being drawn from random chance).  The distributions of the two SLSN populations are also significantly different from each other ($3\times10^{-3}$ chance significance).

More interesting than the simple fact that the SLSN-I host population is skewed towards low-mass galaxies (which has also been noted by others; e.g.,~\citealt{Lunnan+2014,Leloudas+2015}) is the nature of this skew: i.e., is it a preference for the lowest-mass galaxies, an aversion against the highest-mass galaxies, or some combination?  The ratio of the SLSN vs.\ SFR histograms is plotted in the bottom panel of Figure \ref{fig:massrate}: this shows the SLSN rates \emph{per unit star formation}, sometimes called the SLSN production ``efficiency''.\footnote{Mathematically, the SLSN production efficiency $\epsilon$ can be defined as the number of SLSNe per unit time occurring in a cosmic volume divided by the SFR in the same volume: $\epsilon = R_{\rm SN} / SFR$.  This can can refer to all galaxies within a true physical volume, but more frequently is subdivided based on host parameters of interest.  In this case we subdivide it by stellar mass: $\epsilon(M*) = R_{\rm SN}(M*) / SFR(M*)$.  Its absolute units are ${\rm M}_\odot^{-1}$, but because the absolute SLSN rate is quite uncertain, we generally only deal with the \emph{normalized} SLSN efficiency $\hat{\epsilon}$ in this paper; this is the efficiency in a particular stellar-mass bin \emph{divided by} the average cosmic efficiency across all galaxies at similar redshift: $\hat{\epsilon}(M*) = \epsilon(M*)/\epsilon$.  This is a unitless quantity and is what we have plotted in the bottom panel of Figure \ref{fig:massrate}.}
Uncertainties are calculated using the method of \citealt{Cameron2011} (10--90\% confidence interval shown).   A flat value of 1.0 would indicate a perfectly SF-tracing population.  

\begin{figure*}
\centerline{
\includegraphics[width=17cm,angle=0]{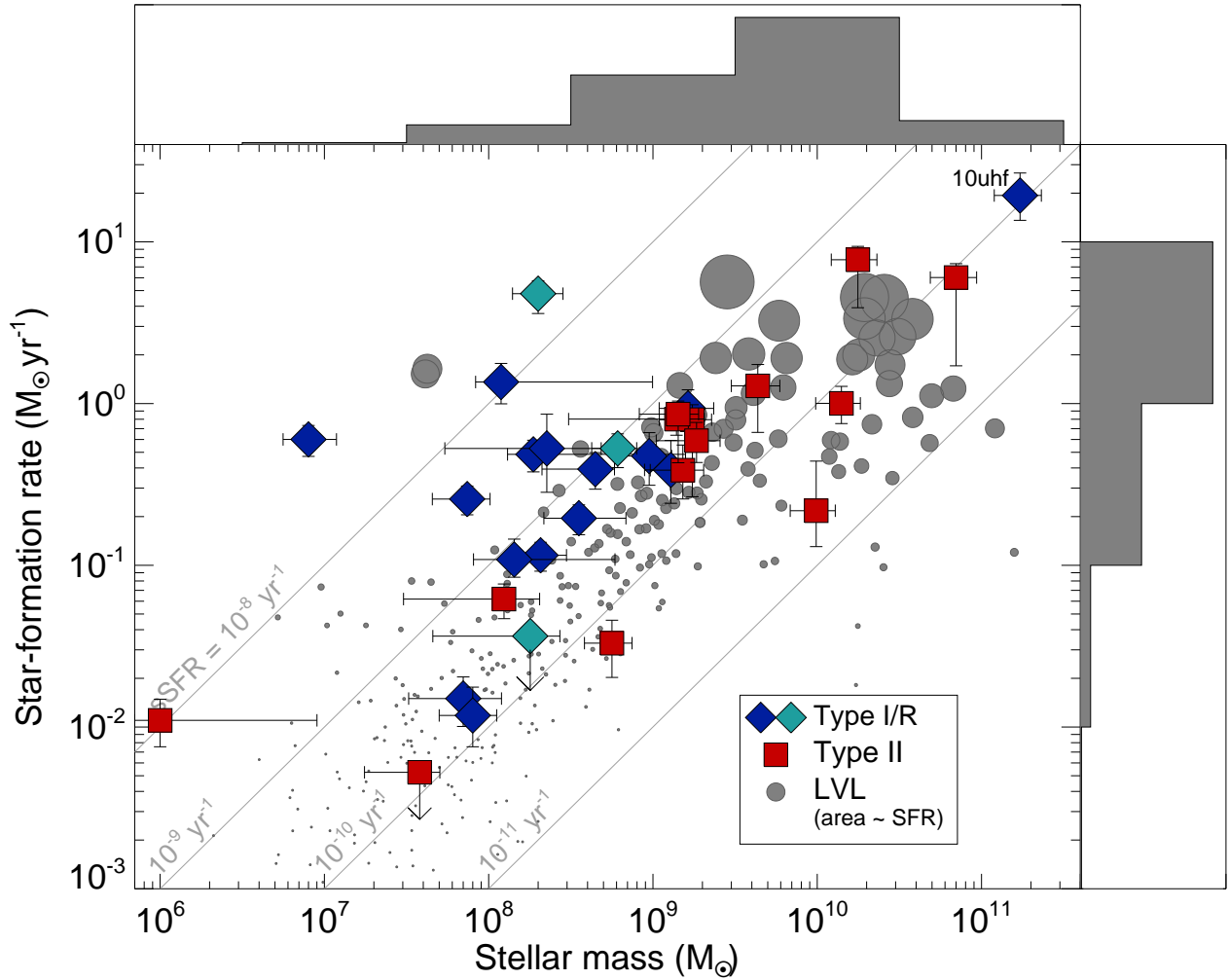}} 
\caption{Star-formation rates and stellar masses of SLSN host galaxies from PTF compared to local galaxies from LVL.  We employ the dust-corrected H$\alpha$ SFRs where this line is detected, or otherwise fall back on the SED-derived average SFRs.  Most galaxies (including most SLSN host galaxies) lie along the star-forming galaxy ``main sequence'' with a specific SFR between $10^{-10}$ and $10^{-9}$\,yr$^{-1}$.  SLSNe-I (but not SLSNe-II) are frequently found in galaxies that are forming stars much more rapidly than average for their mass (``starbursts''), although the majority are on the main sequence.  The histograms on the top and right show the relative contributions of galaxies in LVL to the local SFR density as a function of the parameter on each axis (similar to the middle panel of Figure \ref{fig:massrate}).  Gray diagonals show lines of constant specific SFR.}
\label{fig:masssfr}
\end{figure*}

The Type II SN production efficiency varies by no more than a factor of a few between galaxies of different mass ranges, none of which are significantly different from the uniform null hypothesis (relative efficiency of 1 in all bins), with the exception of the lowest-mass bin.  Because the comparison survey is incomplete at the lowest stellar masses and our number statistics are limited, deviation in a single bin should not be overinterpreted, but it is intriguing that the two faintest galaxies in our sample are both hosts of Type II SLSNe (the host of Type II SN\,2008es is similarly faint; \citealt{Angus+2016})

In contrast, the variation of the type I SLSN production efficiency is enormous between galaxies of different masses, and the bulk of this variation occurs at moderate stellar masses.  While the efficiency varies little between $10^{6}\, {\rm M}_\odot$ to $10^{9}\, {\rm M}_\odot$, it is a factor of 20 less than this in galaxies in the range of $10^{10}-10^{11}\, {\rm M}_\odot$.  (Curiously, it then recovers in the $>10^{11}\, {\rm M}_\odot$ bin, although this determination is based on only a single outlier event, PTF\,10uhf.)

To better quantify this effect, we hypothesized that the SLSN-I rate is constant in galaxies with stellar mass below some value $M_{\rm thresh}$ and nearly zero above the same value.  To test this hypothesis and measure this value, we truncate the Ultra-VISTA sample above $M_{\rm thresh}$ and fit the truncated sample against our host mass distribution to find the optimum value of $M_{\rm thresh}$; we derive a value of $1.6\times10^9\, {\rm M}_\odot$.  (Curves corresponding to this model are shown in light gray in Figure \ref{fig:massrate}).  A K-S test between the truncated galaxy sample and the SLSNe-I in our sample indicates that this model is a good fit ($p=$ 0.38 of being drawn from the same distribution, although the K-S test tends to underweight the impact of extreme values: in this case, PTF\,10uhf.)

These observations seem to argue that the variation of the SLSN-I efficiency is governed by suppression that operates among galaxies with masses above a few $\times 10^{9}\, {\rm M}_\odot$---preventing SLSN formation in these systems (except perhaps in unusual circumstances) but exerting little influence on the efficiency in lower-mass galaxies.

All of these conclusions are, of course, dependent on our sample not being subject to selection effects.  It is possible that the absence of SLSNe in high-mass galaxies may in part be due to biases in targeting transients for spectral classification, although it is very unlikely that the magnitude of the SLSN-I variation can be explained in this way.  This will be discussed in more detail in \S \ref{sec:selectioneffects}.

\subsection{Star-Formation Rates and Specific Star-Formation Rates}
\label{sec:sfr}


\begin{figure*}
\centerline{
\includegraphics[width=17cm,angle=0]{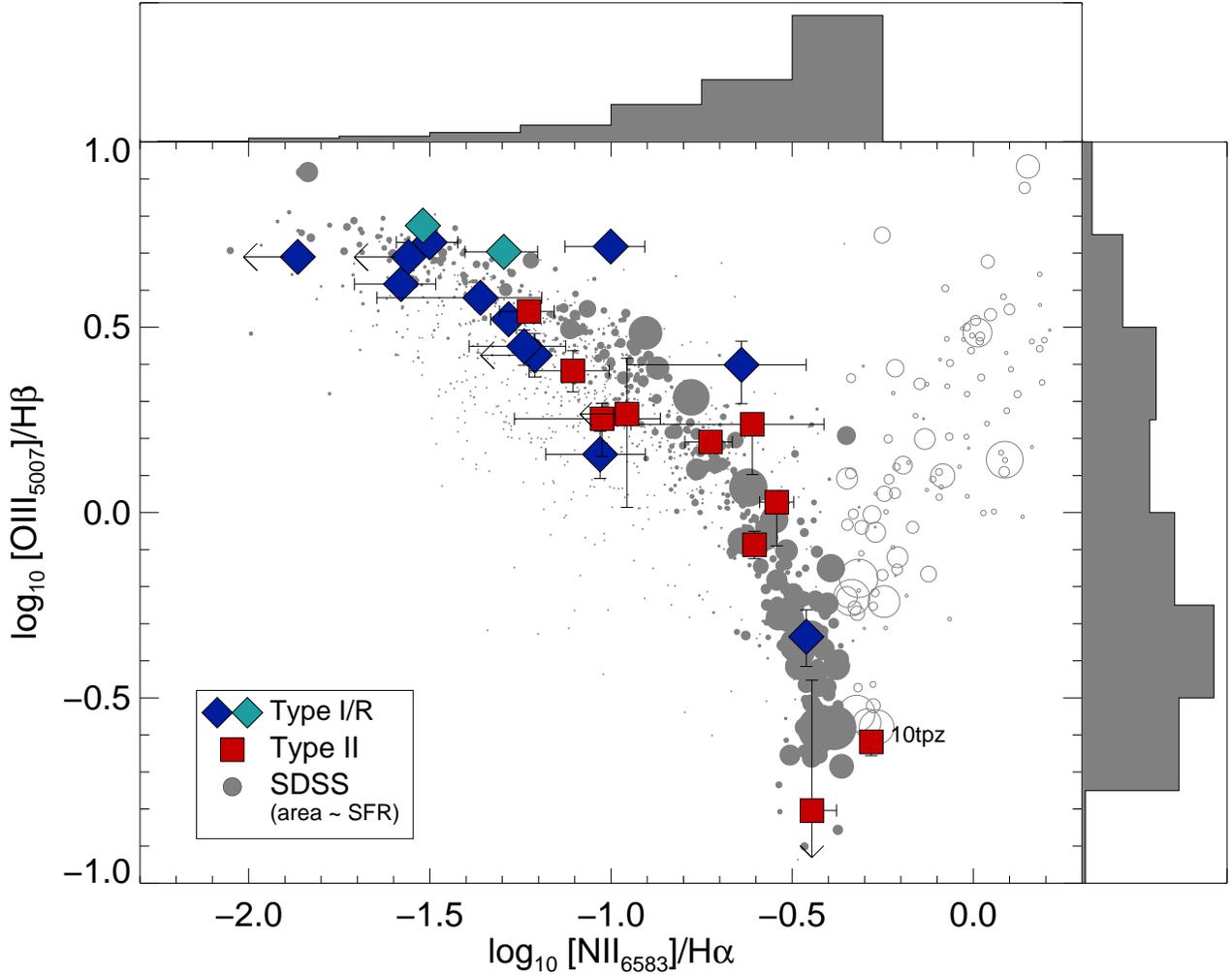}} 
\caption{BPT diagram for SLSN hosts, compared to SDSS galaxies.  AGNs within SDSS are identified via the criterion log([N~II]/H$\alpha$) $> -0.35$ and are indicated as hollow circles.  All SLSN hosts lie on the star-forming branch, and closely follow the same narrow locus as SDSS star-forming galaxies (the host of PTF\,10tpz shows evidence for an AGN contribution to the nuclear flux).  Histograms along the $x$- and $y$-axes show the relative contribution from SDSS galaxies to the SFR density as a function of each parameter (as in Figure \ref{fig:masssfr}).}
\label{fig:bpt}
\end{figure*}

The sample displays a large range of star-formation activity, ranging from $\sim20\,{\rm M}_\odot$\,yr$^{-1}$ all the way down to systems too faint for their star formation to be detected spectroscopically (a limit of typically $<0.1\, {\rm M}_\odot$\,yr$^{-1}$; modeling of the SEDs suggest that the true values are $\sim10^{-2}\,{\rm M}_\odot$\,yr$^{-1}$).  SFR is strongly correlated with stellar mass in typical galaxies (with more massive galaxies producing proportionally more star formation), so the star-formation properties of the sample are best interpreted by plotting SFR together with stellar mass (Figure \ref{fig:masssfr}).  On this diagram we have also marked lines of constant \emph{specific} SFR as diagonal lines.

Most of cosmic star formation, and indeed most SLSNe, occur in a relatively narrow region of the diagram with a specific SFR (sSFR $\equiv$ SFR/$M_*$), sSFR = $10^{-9}-10^{-10}$\,yr$^{-1}$; this feature is a direct manifestation of the $M_*$-SFR correlation and is often referred to as the galaxy main sequence.  We refer to galaxies significantly above the main sequence as ``starbursts'' since their rarity and mass-doubling/gas-consumption timescales imply that these are transient and short-lived phenomena; galaxies well below the main sequence are denoted as ``quenched.''

The high-mass portion of the main sequence is somewhat underpopulated by SLSNe (especially SLSNe-I), as we have already noted in the previous section.  The most interesting feature of this plot, however, is the presence of a significant fraction of our SLSNe-I in extremely active starburst galaxies with SFRs 1--2 orders of magnitude in excess of the $M_*$-SFR correlation.  Depending on the exact stellar masses of these objects (which can be difficult to measure because their light is dominated by the youngest stars), 3--6 Type I SLSN hosts (out of our sample of 18) and 0--2 Type II SLSN hosts (out of 13) are starbursts with sSFR $> 2\times 10^{-9}$\,yr$^{-1}$. 

Galaxies of this type are rare cosmologically.  Their true abundance (and contribution to the local SFR density) is difficult to estimate precisely, but (for example) only 5\% of the star formation in the LVL sample occurs in galaxies with sSFR $> 2\times 10^{-9}$\,yr$^{-1}$.  (See also \citealt{Lee+2009}; similar values are reported from other surveys; e.g., \citealt{Bergvall+2016}).  Given the small sample sizes involved (in particular within LVL: the starburst-galaxy contribution to star formation within the local volume is driven by only two galaxies, NGC1569 and II Zw 40) it is not clear that this trend is statistically significant and we cannot reliably provide an estimate for its quantitative magnitude in the way that we were able to measure the host-mass dependency of the SLSN rates.  However, a similarly high or even higher starburst fraction was seen in the sample of \cite{Leloudas+2015}, so it is unlikely to be a statistical fluke of our sample; furthermore, it is \emph{not} seen in ordinary SNe or even GRBs (\citealt{Sanders+2012,Leloudas+2011,Leloudas+2015}), suggesting it is not a limitation of the LVL volume or the result of redshift evolution.  

\begin{figure*}
\centerline{
\includegraphics[width=17cm,angle=0]{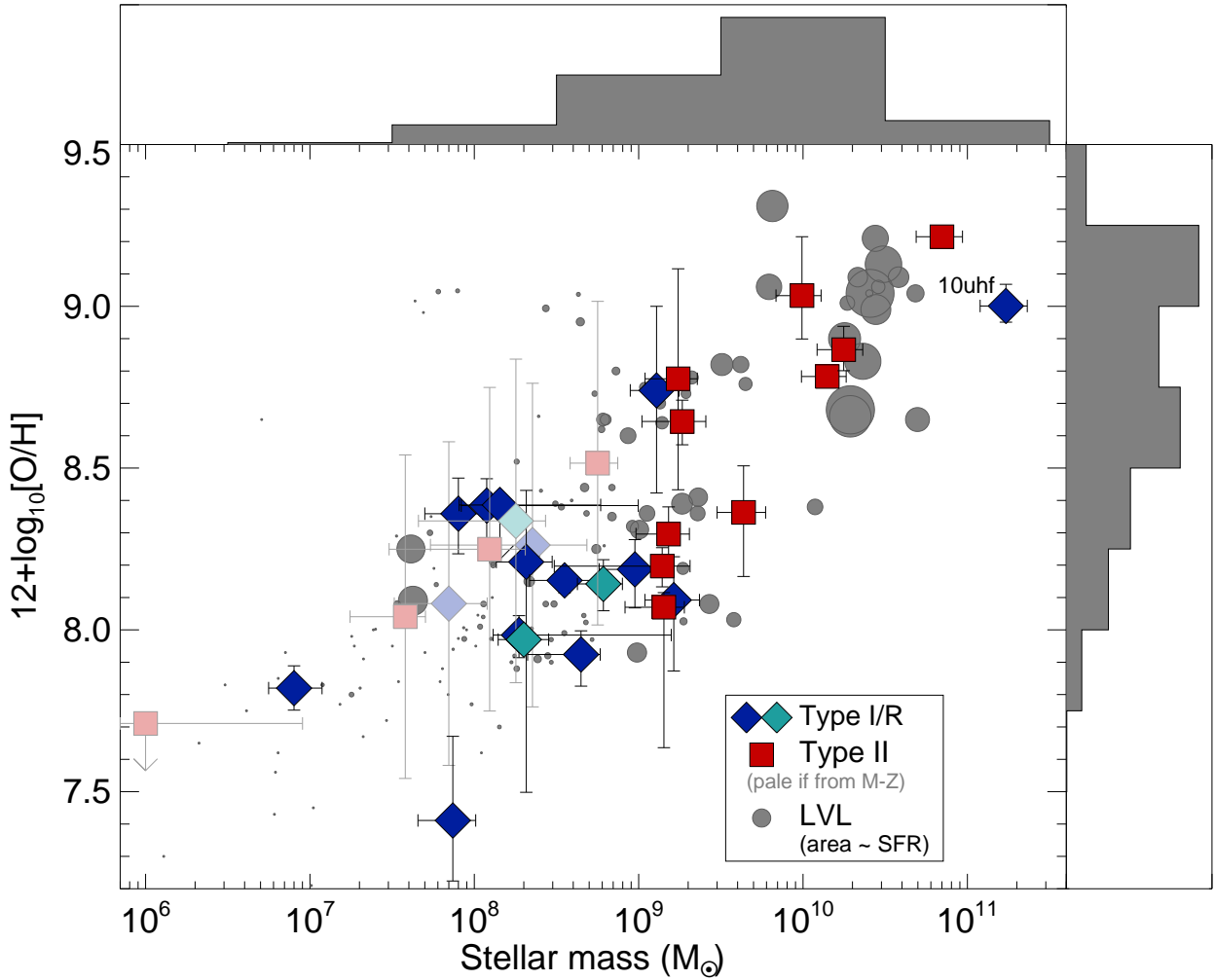}} 
\caption{Mass-metallicity relation for SLSN hosts compared to LVL galaxies.  SLSN hosts generally follow the mass-metallicity relation, allowing for the large intrinsic scatter in this relation (shaded points have metallicities derived from the $M-Z$ relation itself: either because no spectroscopic data was available, or because the emission lines were too weak to permit direct analysis).  SLSNe-I tend to inhabit galaxies with moderately low masses ($\sim10^{8.5}\,{\rm M}_\odot$) and metallicities (12 + log$_{10}$[O/H] $\approx 8.2$).  Histograms are as in Figure \ref{fig:masssfr}.
}
\label{fig:massmetallicity}
\end{figure*}

This preference could, in principle, have the same physical origin as the preference towards low-mass galaxies: possibly, SLSNe prefer starburst galaxies simply because low-mass galaxies are more likely to be starbursts.  (The reverse case can also be considered: SLSNe may prefer low-mass galaxies because only low-mass galaxies undergo starbursts at low-$z$.)   Distinguishing cause from effect in this case is difficult and requires, in particular, an accurate assessment of the starburst contribution to stellar-mass buildup in dwarf galaxies, which is quite uncertain \citepeg{McQuinn+2010}.   We will return to this question later in the discussion (\S \ref{sec:driver}). 


\subsection{Line Ratios}
\label{sec:lineratios}

The BPT \citep{Baldwin+1981} diagram for our SLSN hosts and an SDSS comparison sample (\S \ref{sec:sdss}) is shown in Figure \ref{fig:bpt}.  The usual bifurcation of star-forming and AGN branches is evident; all SLSN hosts lie on the star-forming branch (PTF\,10tpz shows evidence for a subdominant contribution from an AGN).  Indeed, the positions of the SLSN hosts are fully consistent with the locus of low-$z$ galaxies; possible exceptions (PTF\,10hgi and PTF\,10vqv) differ only by $\sim2\sigma$.  They therefore show no evidence of physically unusual environmental conditions (e.g., high-$T$ stellar ionization field or unusual N/O ratio) relative to other low-$z$ galaxies; in particular they do not show the same offsets from the BPT diagram that have been inferred in high-redshift star-forming galaxies \citepeg{Sanders+2015}.  Nevertheless, as can be easily seen from the histogram subplots, there is favoritism for SLSNe towards the top left of the star-forming locus---the region dominated by low-metallicity galaxies.

\subsection{Metallicity}
\label{sec:metallicity}

In Figure \ref{fig:massmetallicity} we plot the stellar mass versus our ``best'' estimate of the metallicity.  Where [O~III] $\lambda{4363}$ is well detected, we employ $T_e$-based metallicities; if it is poorly detected or absent we employ [N~II]/H$\alpha$ using the calibration of \citealt{Nagao+2006}. (We emphasize, however that our results do not depend significantly on the choice of calibration and similar results would be achieved using other diagnostics, and a table of metallicity diagnostics in various alternate calibration systems is presented in Table \ref{tab:linediagnostics}.)  If even H$\alpha$ is absent, we fall back on the metallicity predicted by the redshift-dependent mass-metallicity relation of \cite{Zahid+2014}, adjusted uniformly by $-0.15$ dex to better match the observed relation in LVL; we assume a large uncertainty of 0.5 dex on these values.\footnote{While obviously the use of $M-Z$ is nonoptimal and would be circular for evaluating the relative positions of SLSN hosts relative to the $M-Z$ diagram, excluding these galaxies completely would bias our results since the galaxies with no $H_\alpha$ preferentially are low-mass and low-sSFR.  Only seven of our metallicity values rely on the $M-Z$ relation.}


While the vast spectroscopic coverage and high-quality data provided by SDSS would potentially make this survey ideal as a metallicity comparison sample in many ways, the fiber bias (\S \ref{sec:sdss}), and possible star/galaxy classification biases, introduce severe systematic problems.  Accordingly we fall back on the (small) LVL sample.  While not every galaxy in this sample has a measured metallicity, the subset of galaxies with measurements appears to be representative of the sample as a whole: no biases are evident in mass, SFR, or other parameters.  We caution that the metallicity measurements of this sample (as compiled by \citealt{Cook+2014}) are compiled from the literature and use a variety of different diagnostics: typically $T_e$ at low metallicities and various strong-line diagnostics at higher metallicities.  While this general approach is the same that we employ for our SLSN host sample, the actual diagnostics employed are not always the same, introducing additional systematic uncertainty in our comparisons that should be kept in mind during the ensuing discussion.

Consistent with previous studies \citep{Lunnan+2014,Chen+2015}, we find that SLSNe-I are highly underabundant in the most metal-rich galaxies.  All but two occupy galaxies with metallicities\footnote{These numbers refer to our ``preferred'' values combining the $T_e$ and N06 metallicity scales, but equivalent results are obtained if any major scale is adopted.  For example, if the KK04 scale is exclusively adopted, metallicities increase by about 0.1 dex in most of our galaxies but all targets except for the host of PTF\,10uhf remain below 12 + log$_{10}$[O/H] $< 8.55$.} below 12 + log$_{10}$[O/H] $\lesssim 8.5$, even though the majority of local star formation occurs at higher metallicities.  Their metallicities, however, tend to not be \emph{much} lower: all but one have values above 7.8 and the median value is 8.2 (1/3 Solar).  In addition, with or without the mass-metallicity-based values included, the position of the SLSN hosts in Figure \ref{fig:massmetallicity} gives no indication of an offset relative to the local mass-metallicity relation.  This suggests that the metallicity dependence of the SLSN-I production efficiency follows similar qualitative behavior as the stellar-mass dependence: uniform at values below a critical threshold, above which the production efficiency drops precipitiously.  

Given the difficulty of establishing a complete comparison sample of star-forming galaxies using identical metallicity-measurement techniques as our host sample, it is difficult to measure this threshold precisely, but it must lie above the majority of our SLSN host metallicities (12 + log$_{10}$[O/H] $\gtrsim 8.2$--8.4) yet below the metallicities of ``typical'' spirals, which dominate the cosmic SFR density and are almost never found to host SLSNe (12 + log$_{10}$[O/H] $\lesssim 8.6$).  A reasonable working hypothesis---to be tested and refined by future studies\footnote{We note that, following the initial submission of this paper, a 0.5\,$Z_\odot$ metallicity cutoff for SLSN-I production was independently proposed by \cite{Chen+2016}.}---is that the SLSN efficiency is constant below half-Solar (12 + log$_{10}$[O/H] $< 8.4$) and extremely low above this value.

\section{Discussion}
\label{sec:discussion}

\subsection{Selection Biases?}
\label{sec:selectioneffects}

It is of obvious importance to consider to what extent any of the results above might be influenced by selection effects.  Biases can be intrinsic (due to extinction or source confusion/blending preventing detection in the imaging survey) or extrinsic (due to follow-up/classification biases preventing recognition that some events present in the imaging survey are indeed SLSNe).

\subsubsection{Extinction and Confusion}

As SLSNe are always optically selected, extinction-related bias is unavoidable.  It is probably not, however, particularly important in the case of our sample.  First, most low-redshift galaxies are optically thin ($A_V<1$ mag, as confirmed by the LVL extinction data); most star formation is not heavily obscured.  Indeed, most optical surveys---including PTF itself---have no difficulty finding other types of SNe in massive galaxies (see, e.g., \citealt{Arcavi+2010} for the luminosity distribution of the PTF core-collapse sample, and our discussion in the following section).  Furthermore, if extinction were a major factor, we would expect its impact to be strongly dependent on redshift: SLSNe at $z\approx0.1$ could be found relatively easily even behind 1--2 mag of extinction, whereas those at $z\approx0.4$ would be affected most strongly.  However, the lack of massive galaxies in the SLSN sample is evident at low redshifts as well as at high redshifts (Figure \ref{fig:z_luminosity}).  Furthermore, an extinction effect would not explain the apparent differences between the host populations of SLSNe-I and SLSNe-II.

Source confusion is even less likely to be an issue, at least for the sample as a whole: PTF subtractions are generally quite clean except around the centers of massive nearby galaxies (which contribute little to star formation).  Confusion-related biases could, however, conceal a population that occurs preferentially or exclusively in these environments.  This may indeed be relevant in the case of PTF\,10tpz, but it has no impact on the remainder of our sample.

\subsubsection{Spectroscopic Follow-up Biases}

A much more significant consideration is related to the fact that only a minority of PTF transients can be followed-up and classified.  While careful reanalysis of archival data suggests that the identification of transient candidates by scanners is highly complete and not biased (Frohmaier et al., in prep.), the manner in which candidates are chosen for spectroscopic follow-up is not uniform and heavily depends on human-based decision-making using information available around the time of discovery (\S \ref{sec:classification}).  These decisions may include the apparent properties of the host in pre-imaging in some cases.   Indeed, transients in faint/dwarf-like hosts and transients with no apparent host in pre-imaging are often explicitly flagged as interesting relative to transients within large and resolved intermediate-redshift ($z\approx0.1$--0.2) galaxies, since the large majority of transients within the latter type of galaxy are Type Ia SNe (which, because of their numbers, are not usually emphasized for follow-up unless they are discovered very early or occur very nearby).  

An approximate accounting for this effect can be measured by examining the host-galaxy sample for general PTF core-collapse SNe compiled by \cite{Arcavi+2010}.  These SN hosts exhibit a median absolute magnitude of $M_r=-20$ with only a small fraction (19\%) found within ``dwarf'' ($<-18$ mag) galaxies.  This is, in fact, quite similar to the fraction of cosmic SFR occurring within galaxies of this absolute magnitude range (23\%, for Ultra-VISTA galaxies at $z\approx0.12$) and suggests that, at least in the first few years of PTF, galaxy-related classification biases did not have a large impact.   

\begin{figure*}
\centerline{
\includegraphics[width=17cm,angle=0]{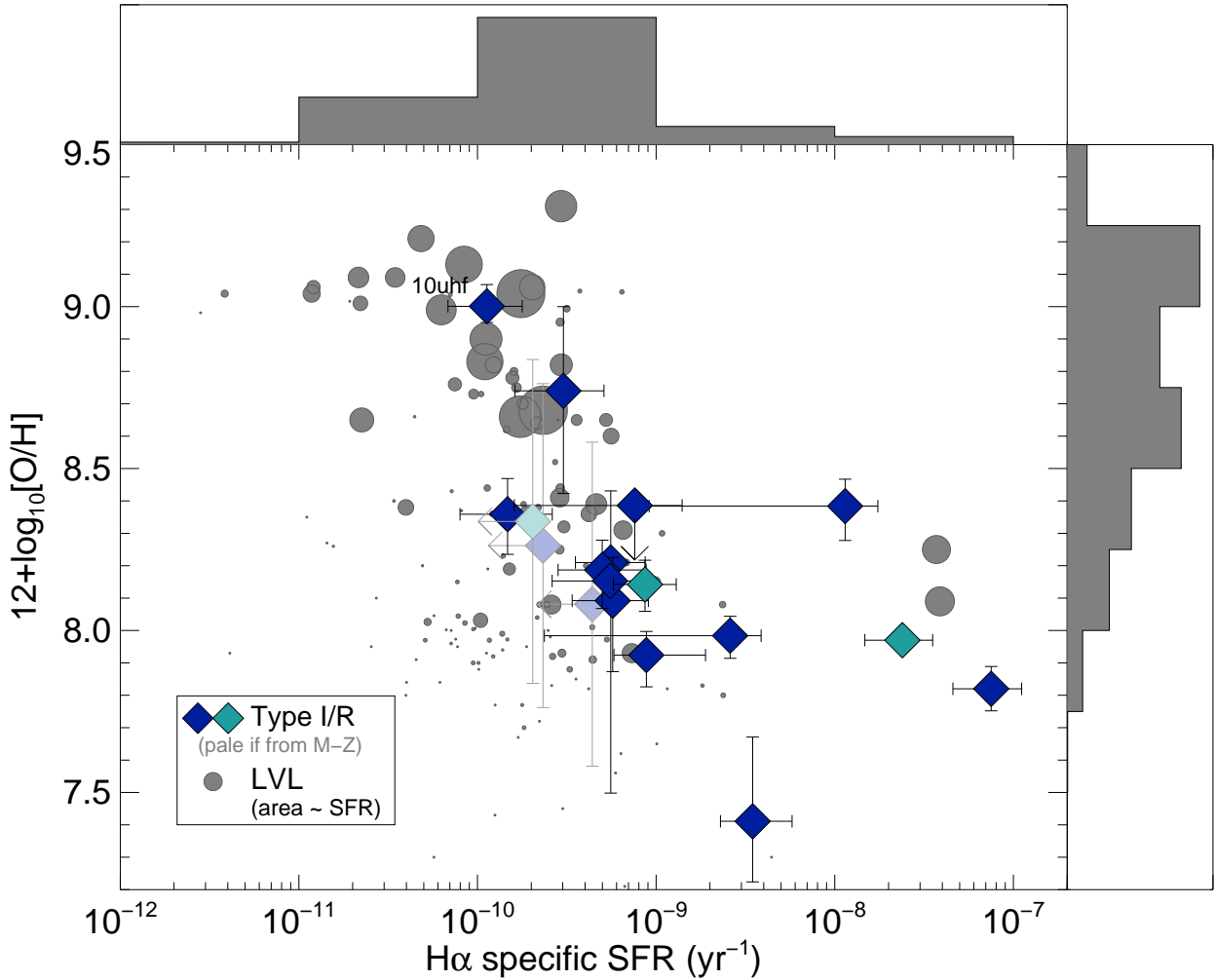}} 
\caption{Specific star-formation rate versus metallicity.  This plot can be used to help distinguish whether metallicity or star-formation intensity is primarily responsible for the unusual properties of the SLSN-I host population: a dominant metallicity effect would be seen as a concentration in the vertical direction relative to LVL galaxies, while a dominant SFR-intensity effect would be seen as a concentration in the horizontal direction.   Many of the sSFR measurements are upper limits due to nondetection of H$\alpha$.  SLSNe-I clearly avoid metal-rich galaxies, and may prefer somewhat elevated sSFRs compared to the LVL sample but do not require it.  The three ``extreme'' SLSN-I hosts in the right part of the figure (PTF\,10vqv, PTF\,11dij, and PTF\,12dam) have few equivalents among local-volume galaxies.  Histograms are as in Figure \ref{fig:masssfr}.
}
\label{fig:ssfrmetallicity}
\end{figure*}

In addition to possible biases in \emph{obtaining} spectroscopic follow-up, there may be biases in \emph{interpreting} spectroscopic follow-up if transients within certain types of galaxies are easier to classify.  For example, faint galaxies might contaminate the transient spectra less and make their features easier to recognize (this would lead to a bias against luminous galaxies), or underlying emission lines might reveal the redshift and therefore reveal that the transient is ``superluminous'' more readily (this would lead to a bias in favor of galaxies with strong emission lines).

We do not think these possible biases are significant in our sample.  SLSNe are comparable in brightness to even the most luminous galaxies and it is very difficult to conceal their intrinsic features regardless of galaxy luminosity or type.  In addition, at all but the lowest redshifts the UV absorption features that emerge in SLSN-I post-peak are very clear and distinctive; furthermore a careful late-time reanalysis of the entire PTF spectral database (Quimby et al., in prep.)\ identified only a single misclassified SLSN-I, and we have included this event (PTF\,09as) in our sample.   The redshifts of SLSNe-II are even more immediately apparent at the time of the first spectrum thanks to their narrow Balmer-series emission lines, although on the other hand it is conceivable that some SLSNe-II with spectra might be missing from our sample due to misclassification as luminous AGNs/QSOs (which could bias SLSNe-II against pointlike galaxies or small nuclear offsets).

Based on these arguments, we conclude that the trends evident in the PTF SLSN-I sample above are genuine and reflect a true, intrinsic increase in likelihood of these events in low-mass, low-metallicity, high-sSFR galaxies.   In the case of SLSNe-II, our conclusion that they differ from the properties of general star-forming galaxies was only weakly significant in the first place, and it is conceivable that this trend could be erased once some of the selection effects above are considered.  For these reasons, in the rest of the discussion we will focus primarily on the SLSN-I population.

\subsection{What Drives the SLSN-I Rate Variation: Metallicity, Starbursts, Both, or Neither?}
\label{sec:driver}

We previously (\S \ref{sec:results}) noted two strong trends evident in the SLSN-I host population.  First, SLSNe-I are very uncommon in galaxies with high masses and metallicities ($M_*>10^{9.5}\, {\rm M}_\odot$ and $Z>0.5\,{\rm Z}_\odot$, respectively)---which could be interpreted as evidence for a metallicity-dependent progenitor mechanism.  Second, they are remarkably common in ``starburst'' galaxies with high specific SFRs (sSFR $> 10^{-9}$\,yr$^{-1}$)---which could be interpreted as evidence for a progenitor favored by an altered (e.g., top-heavy) IMF or dense clusters with abundant dynamical interactions.   Is this evidence that \emph{both} metallicity \emph{and} an IMF/interactions contribute to producing SLSN progenitors?
 
Caution is warranted on this point:
low-mass(/metallicity) galaxies have burstier star-formation histories than high-mass(/metallicity) galaxies \citepeg{Lee+2009} and when starbursts do occur they can be much more intense in dwarf galaxies.  Furthermore, galaxies mid-starburst may exhibit lower gas-phase metallicities even when compared to others of the same mass \citep{Mannucci+2010}.  Given these influences, the simultaneous observation of both of the \emph{observational} trends we highlight above is to some degree expected \emph{even if} only one effect is operating---simply because the parameters correlate.

Because of the much greater strength of the metallicity trend (most star formation is metal rich, most of our hosts are metal poor) compared to the sSFR trend (most star formation is in nonstarbursts, and indeed most of our hosts are also nonstarbursts) it is more likely that a metallicity effect would be driving the starburst trend rather than vice versa.  However, we can test this more directly by taking advantage of the fact that the correlations between galaxy parameters are not perfect ones, and examine the role of both factors simultaneously in a two-dimensional parameter space.

In Figure \ref{fig:ssfrmetallicity} we plot the two competing parameters ($Z$ and sSFR) against each other directly, along with LVL galaxies with measured metallicities.  Type II SLSNe are omitted for clarity.  If metallicity is the dominant effect driving SLSN production the hosts would preferentially avoid the top part of the diagram and cluster towards the middle and bottom, but their horizontal distribution would match local star formation within that space.  If starburst intensity is a critical factor, SLSNe would preferentially occur towards the right side of the diagram relative to star-forming galaxies of the same metallicity.

We can rule out that the starburst trend \emph{drives} the metallicity trend: even if starburst galaxies are removed, most of the hosts remain quite metal-poor compared to the majority of star formation in the local volume.   It is less clear whether the reverse is also true (that is, whether a preference for starbursts remains if only the most metal-poor galaxies are considered).  The three SLSN host galaxies that are strongly starbursting (PTF\,10vqv, PTF\,11dij, PTF\,12dam) are also definitively metal-poor.  In our local-volume comparison sample, the two most dramatic local-volume starbursts have comparable sSFRs to our three SLSN starburst hosts, but all three SLSNe-I starbursts are poorer in metals---providing at least some suggestion that metallicity is the driving factor even in the high-sSFR regime.  However with sample sizes of only three hosts and two comparison galaxies this statement has almost no statistical significance and further studies of larger samples will be needed to evaluate this more definitively.

An alternative means of addressing this point would be to examine the precise spatial positions of the SLSNe \emph{within} their hosts, especially for the starburst galaxies, to determine whether or not the SLSNe are indeed occuring in regions of the most intense star formation.  This has been done previously by \cite{Lunnan+2015} on a higher-redshift sample; they find that SLSNe may slightly concentrate towards the brightest pixels but that this trend is not statistically significant.  One of their events is a starburst galaxy in our sample (PTF\,11dij/SN\,2011ke) and the SN appears on a pixel with a UV surface brightness higher than 86\% of the remainder of the galaxy.  In our own imaging of PTF\,12dam, the SLSN sits on the brightest pixel of the F225W ultraviolet image (De Cia et al., in prep.).  This may suggest that \emph{very} high star-formation volume densities may amplify the SLSN rate independent of metallicity, but that this effect drops away at ordinary SFR densities.  We will need more UV imaging of the starburst subset of the SLSN-I host population to confirm this possibility.

\subsection{The Peculiar Hosts of PTF\,10uhf and PTF\,10tpz}
\label{sec:10uhf10tpz}

The host galaxy of PTF\,10uhf stands as a stark exception to many of the trends seen in the rest of the SLSN-I host sample.  It is massive, metal-rich, and has a large population of evolved stars.  Because the SN occurred far in the outer regions of this galaxy (possibly in a merging, dwarf companion), it is natural to appeal to the possibility of metallicity gradients to explain this peculiar case under the ``maximum-metallicity'' hypothesis we propose (\S \ref{sec:metallicity}).  However, analysis of spectroscopy taken at the SN site shows no significant difference between this region and the rest of the host: the metallicity at the SN site is still well above our proposed threshold.  Possibly, small-spatial-scale chemical inhomogeneities \citep{Niino+2011} could conceal some low-metallicity star formation, although given the weakness of the [O~III] line this would have to represent a small fraction of the overall total.  Alternatively, this host may provide evidence that SLSNe-I are not completely suppressed in metal-rich environments: they may still occur at high-$Z$, but at a much lower frequency.  

Curiously, however, this host is at the extremes not only of the SLSN host population but also of the general galaxy population.  If in principle a SLSN-I can occur in any star-forming galaxy (due to chemical inhomogeneities, incomplete suppression at high metallicity, or other reasons) it is peculiar that the only metal-rich SLSN-I in our sample happened to occur in one of the most luminous galaxies in the low-redshift universe.   We might wonder, then, whether something exceptional about this galaxy might have enabled it to produce this event in a metallicity regime which would stifle the production of SLSNe in most other galaxies: perhaps the metallicity limit might be ``lifted'' in regions of the most vigorous star formation owing to an altered IMF.  This specific explanation is not particularly natural either: star formation at the SN site, while clearly present, is not particularly vigorous (in terms of, e.g., [O~III]/[O~II] or H$\alpha$ equivalent width).  However, it is possible that there are other unusual physical attributes of the host and/or the SN site that are less apparent.

With only a single example it is not possible to rule out any of these scenarios: the seemingly exceptional nature of its host could simply be a low-probability event, and future ``high-metallicity'' SLSNe will be found in more normal galaxies.  Selection biases within PTF, while minor (\S \ref{sec:selectioneffects}), may also play a role.   In any case, the occurrence of rare but dramatic outliers of this type clearly indicates the need to continue collecting large samples of SLSNe and their hosts to determine just how frequently the cosmos sees fit to violate the ``rules'' we might otherwise infer from smaller samples.  

Indeed, our discovery of PTF\,10tpz (a Type II SLSN) provides a possible illustration, showing that the very peculiar environment of SN\,2006gy (the circumnuclear regions of a massive and metal-rich galaxy) was probably \emph{not} a fluke.  Heavy obscuration and confusion limit the effective search volume for heavily obscured circumnuclear events of this type within PTF to a volume much smaller (by $\gtrsim$2 orders of magnitude) than for unobscured SNe.  Accounting for this, obscured circumnuclear SLSNe may in fact constitute the most common variety of SLSN.  This is especially notable considering that the contribution to cosmic star formation from massive galaxy nuclei is miniscule.  Low metallicity clearly is not the factor in this case, suggesting that---in the case of circumnuclear Type II SLSNe---a genuinely different factor is at play, such as a top-heavy IMF in galaxy nuclei. The presence of a central AGN in the host galaxies of both SN\,2006gy and PTF\,10tpz provides some evidence that the AGN itself may play a major role in shaping the nuclear star-forming environment.

Alternatively, it may be worth considering whether---despite being clearly located away from their host nuclei and exhibiting SN-like photometric and spectroscopic evolution---SN\,2006gy and PTF\,10tpz may not represent genuine SNe after all, but rather flares associated with a \emph{second} supermassive black hole originating from a former companion galaxy that merged with the host in the recent past.  The host of PTF\,10tpz does indeed show some suggestions of a past merger in the form of two nonaligned disks (Figure \ref{fig:mosaic}).  A detailed examination of this hypothesis will be reserved for future work.

\subsection{Implications of a Metallicity Threshold for the SLSN-I Progenitor}

The roughly half-Solar metallicity threshold for SLSN-I production that we propose\footnote{As we measure only galaxy-averaged metallicities, it is possible that the true metallicities at the progenitor site are somewhat lower than our reported values \citep{Niino+2011}.  However they are unlikely to be dramatically lower as an ensemble, especially considering the small physical sizes of most of our objects.} poses a broad challenge to theoretical models for SLSN production.  While a SLSN rate that depends strongly on metallicity is a natural prediction under a variety of scenarios, most of these suggest a critical threshold that is much lower ($\sim0.1\,{\rm Z}_\odot$, or even less)---or, at least, a smoother dependence in which the rate at very low metallicities is significiantly higher than the rate at moderate metallicities.

Classical pair-instability (PI) models, including the pulsational pair-instability variant, suggested that only extremely low-metallicity stars should explode as PI-SNe \citepeg{Heger+2003,Langer+2007}.  Our observations would therefore seem to rule out this model for SLSNe-I, including the slowly declining ``R'' subclass.  However, more recent theoretical work suggests that the picture is more equivocal: for example, \cite{Yusof+2013} are able to produce pair-instability explosions at up to $Z\approx0.5\,{\rm Z}_\odot$ (see also \citealt{Marchant+2016}).  However, these studies also predict that the rate at very low metallicities should be much higher than at $Z\sim0.5Z_\odot$, contrary to our observations.  While individual SLSNe in our sample may remain viably explained by PI-SNe, it remains unclear why we do not find more examples in galaxies that are (even) lower in mass and metallicity than those in our sample.

Explaining a relatively high metallicity ``threshold'' in the SLSN-I efficiency is challenging for other models as well, although the wider range of evolutionary pathways and initial masses available to both the magnetar model and core-collapse+interaction model (especially when binary evolution is considered) may offer some additional flexibility.  Interestingly, this problem is shared with GRBs, whose rate dependence on metallicity seems to exhibit at least qualitatively similar behavior as what we report here for SLSNe \citep{Kruehler+2015,Vergani+2015,Perley+2016b}.

Whatever the underlying reason, this similar behavior provides some evidence for a close connection between the progenitors of SLSNe-I and GRBs---and since the progenitor of GRBs is almost certainly \emph{not} super-massive given the much lower typical luminosities and ejecta masses of the SNe accompanying GRBs, this might argue that a central-engine model (which almost certainly explains GRBs) applies for SLSNe-I also \citep{Lunnan+2014}.

On the other hand, the inferred metallicity threshold for GRBs from the studies above ($\sim 1.0\, {\rm Z}_\odot$) is significantly higher than what we measured for SLSNe-I in this work ($\sim 0.5\, {\rm Z}_\odot$).  These studies were conducted at different redshift ranges and (in some cases) using different techniques, and until a large, complete comparison sample of SLSNe and GRBs at similar redshifts is available we should exercise caution in comparing their environmental preferences directly.  Nevertheless, other forms of evidence also suggest that the host environments of GRBs and SLSNe-I do in fact differ (\citealt{Vreeswijk+2014,Leloudas+2015,Angus+2016}, as well as the low-$z$ subset of \citealt{Lunnan+2014}).  Further studies of both populations will be needed to confirm that these differences are intrinsic and significant, and genuine differences between the GRB and SLSN host populations may indeed be present even if they share a common central engine (clearly, the very different nature of these events requires \emph{some} differences in their formation histories).  In any case, we cannot yet argue that the similiarities between SLSN and GRB hosts establish a clear connection between their progenitors or energy sources.

The least exotic progenitor model---an ``ordinary'' but high-mass core-collapse explosion interacting with a massive hydrogen-poor envelope---is perhaps the most difficult to rule out since it requires neither rapid rotation nor an extremely large core mass, establishing fewer constraints on its evolutionary pathway before explosion.

\subsection{Constraints from Young Starburst Ages on the SLSN-I Progenitor Mass}

In cases where a host galaxy is undergoing a particularly young and active starburst, its properties can be used more directly to constrain the nature of the progenitor.  It is reasonable to assume in these cases that the progenitor was formed during the starburst episode and that its lifetime cannot be longer than that of the starburst itself, placing a maximum age on the progenitor that can be translated to a minimum initial mass.  Only quite extreme galaxies are capable of providing a meaningful constraint: any star large enough to explain the ejecta masses inferred from typical SLSNe will have an initial mass well exceeding 20\,M$_\odot$ and a lifetime shorter than 10\,Myr, so galaxies undergoing starbursts older than this provide no additional information about the progenitor age.  Fortunately, a few SLSN hosts are indeed quite extreme.  The host of PTF\,12dam is the best-studied example; a high-S/N spectrum of this galaxy has been previously used by \cite{Thoene+2015} to conclude that the starburst age was $\sim2$\,Myr, corresonding to an initial progenitor mass of $>60\, {\rm M}_\odot$.  The host of PTF\,11dij is even more extreme in its color and equivalent widths and we suggest that more detailed modeling of this event may be able to place even more stringent constraints on the initial mass range of its progenitor.  Very massive progenitors are required in the PI-SN and interacting CC-SN models but lower-mass progenitors are more likely under the magnetar model, so these observations may therefore argue in favor of one of the former two cases.

However, we caution against attempts to generalize these conclusions to the entire sample: while their extreme properties attract attention, hosts like PTF\,12dam and PTF\,11dij represent a small minority of the population, and the sSFRs of most of our hosts (including the other starbursts) correspond to characteristic ages of 100\,Myr or more.   This is not necessarily problematic for an ultra-massive progenitor star (such stars are expected to form outside young starbursts as well).  On the other hand, the resolved {\it HST} analysis of \cite{Lunnan+2015} shows no clear tendency for SLSNe to prefer the most UV-luminous portions within their host galaxies, as would be expected if their progenitor was exclusively very short-lived.  Possibly, SLSNe-I exhibit a range of explosion timescales and therefore initial masses.  More examples of extreme-starburst host-galaxy systems, and a better measurement of their true frequency, will be needed to resolve this.

\subsection{The Absolute Efficiency of SLSNe in the Most Metal-Poor Galaxies}

The \emph{absolute} rate of SLSNe also provides a constraint on progenitor models \citep{Chen+2013}.  For example, exotic scenarios involving a series of low-probability events could be ruled out if they do not produce enough SLSNe in the universe to explain the numbers that we observe. Or, a simple model in which an initial mass above a certain value is the \emph{only} criterion can be ruled out if that value implies the production of too \emph{many} SLSNe compared to the observed rate.  Such constraints can in principle be strengthened by factoring in host-galaxy preferences: since most SLSNe-I originate from a small subset of the overall galaxy population, the fraction of stars exploding as SLSNe in this type of galaxy is significantly higher than the ``cosmic average'' would suggest.

The overall average cosmic rate is constrained only very approximately at present. \cite{Quimby+2013} previously estimated the SLSN-I rate density at $z\approx0.15$ to be roughly within the range of $10^1$--$10^2$ events Gpc$^{-3}$\,yr$^{-1}$, compared to an overall CC-SN rate of $\sim10^5$ at the same redshift \citepeg{Strolger+2015}: i.e., SLSNe-I represent one out of every 10$^3$ to 10$^4$ SNe at this epoch.

This figure averages together the rate from metal-poor dwarf galaxies (which produce almost all of the SLSNe-I) with the massive, metal-rich hosts (which produce almost none of them).  Given the ``step''-like behavior of the SLSN-I rate we infer (as a function of mass and probably of metallicity; Figure \ref{fig:massrate}), it makes sense to calculate separately the fraction in low-mass versus high-mass galaxies.  This can be determined simply by multiplying the ``average'' number above by the values plotted in the bottom panel of Figure \ref{fig:massrate}.   In galaxies with masses below $10^9\, M_\odot$, the fraction is higher by a factor of 5: SLSNe-I constitute one per 200--2000 SNe.  In galaxies with masses above $10^{10}\, M_\odot$, the fraction is instead \emph{lower} by a factor of 5 (or more if 10uhf is ignored):  SLSNe-I constitute one per ($>$)5000--50000 SNe.

Given the rarity of extremely massive stars, a relatively high SLSN rate becomes interesting for constraining high-mass progenitor models.  For example, if the SLSN-I rate is at the high end of the \cite{Quimby+2013} estimate ($10^2$ Gpc$^{-3}$\,yr$^{-1}$), then at $Z<0.5\,{\rm Z}_\odot$ SLSNe-I constitute 0.5\% of all core-collapse SNe, 5\% of SNe from progenitors with $M_{\rm init} > 50\,{\rm M}_\odot$, and $\sim50$\% of SNe from progenitors with $M_{\rm init} > 200\,{\rm M}_\odot$.  (We use a high-end IMF slope of $\alpha=2.3$, and assume that all stars with $M_{\rm init} > 8\,{\rm M}_\odot$ explode as SNe.)  The similarity of the SLSN rate to the high-mass star SN rate is particularly intriguing.  If the intrinsic SLSN rate is at the low end of the Quimby et al.\ estimate, all these rates would drop by a factor of 10 and additional factors beyond mass and metallicity would become necessary to explain the SLSN rate \emph{even if} the progenitor is extremely massive.

\section{Conclusions}
\label{sec:conclusions}

The hosts of SLSNe-I have highly divergent properties from the general star-forming galaxy distribution: lower masses and metallicities (in nearly all cases) and unsually high specific SFRs (in a notable minority of cases).  These trends do not appear to be due to selection effects and suggest a progenitor whose production is intrinsically favored in some environments and suppressed in others.  Most likely, the primary factor influencing this is a requirement of low (but not extreme) metallicity: below a galaxy-averaged oxygen abundance of 12 + log$_{10}$[O/H] $\lesssim 8.4$ (equivalent to 0.5\,Z$_\odot$) the SLSN-I rate rises by approximately a factor of 20 higher compared to galaxies with metallicities higher than this value.  The rates of SLSNe-I in metal-poor galaxies at $z\sim0.2$ are about a factor of 5 higher than what would be implied by their cosmic ``average''.

A metallicity limit alone appears to provide good consistency with all of the bulk properties of the SLSN-I host population without a need to introduce other factors (such as a dependence on specific SFR), and the observed abundance of SLSNe-I in starbursting galaxies may simply reflect the bursty star-formation histories of metal-poor, low-mass ($\sim10^{8}\, {\rm M}_\odot$) galaxies.  This comparison is limited by our small sample size and by the lack of large volume-complete spectroscopic surveys of nearby dwarf galaxies, however.  In any case, the majority of SLSNe-I in our sample occupy host galaxies with SFRs typical of their stellar masses, indicating that the role of an sSFR-dependent IMF or dynamical interactions is at best secondary.

The functional dependence of SLSNe-I on metallicity we prefer is qualitatively similar to what has been inferred by recent work on GRBs: a constant or slowly variable rate up to a threshold, above which it drops sharply.  However, the threshold for SLSNe ($\sim0.5\,{\rm Z}_\odot$) appears to be lower than for GRBs ($\sim1\,{\rm Z}_\odot$), so this similarity does not necessarily argue for a common origin.  In any case, no first-principles progenitor model that we are aware of predicts the type of metallicity dependence that we infer for either transient, and more theoretical work is necessary to explain why the rates of energetic transients at 0.4\,Z$_\odot$ (SLSNe and GRBs common), 0.7\,${\rm Z}_\odot$ (only GRBs common), and 1.5\,Z$_\odot$ (neither event common) appear to be so dramatically different despite differences of factors of only a few in metal abundances.  

While this behavior is theoretically puzzling, observationally it may render the problem more tractable: the relatively high metallicity threshold we infer, and the relatively ``ordinary'' nature of most host galaxies, suggests that the SMC and quite possibly the LMC are quite viable SLSN (and GRB) hosts.  While the possibility of an actual SLSN occurring in one of these galaxies anytime in the near future is of course miniscule, they may contain stars that are SLSN progenitors.  Even if not, resolved-population studies will allow direct investigation into the differences of massive stellar populations over the range of metallicities relevant to controlling SLSN-I production.  Interestingly, there are already some indications from resolved studies that the LMC can form stars at $>200\, {\rm M}_\odot$ but the Milky Way cannot \citep{Crowther+2010,Crowther+2016}.  If the progenitor of SLSNe-I is indeed a very massive ($M_{\rm init}>200\,{\rm M}_\odot$) star, this may indicate the manner by which metallicity affects SLSN production has more to do with massive star \emph{formation} rather than massive star \emph{evolution}.

All of the trends we discuss above are weaker, or absent entirely, among Type II SLSNe.  While we find a modest tendency for SLSNe-II to favor lower-mass hosts compared to cosmic star formation generally, this effect is not strong and some of it may be attributable to selection effects.  A larger Type II host sample will be needed to resolve this question unambiguously, but we suggest that most Type II SLSNe likely represent a rare but not particularly environmentally dependent extreme of the same physical process that generates ``ordinary'' Type IIn SNe.   The circumnuclear SLSNe represented by SN\,2006gy and PTF\,10tpz seem to represent a different situation entirely: these objects likely belong to a distinct class of transient, exclusive to these environments.

Recently, the ASAS-SN team announced the discovery of what they refer to as the ``brightest supernova ever'' \citep{Dong+2016}.  This event is reported to be of Type I, and is located at the centroid of an extremely massive and red galaxy with no star formation present.  Given the host-galaxy properties of Type I SLSNe (and even Type II SLSNe) within our own sample, we are skeptical that this event represents a genuine SLSN and suggest that it is more likely a tidal disruption event due to a star falling into the central black hole (the black hole in this galaxy would have to be unusually small given the galaxy's mass for a TDE to be observable, but this could be the case because of a dual SMBH from a minor merger, for example).  If it is a genuine SN, it would suggest that whatever peculiar factors govern star formation near the centers of massive galaxy nuclei can favor Type I as well as Type II SLSNe.

\vskip 0.02cm

\acknowledgments

We thank the anonymous referee for a careful reading of the submitted paper and for providing valuable comments.   D.A.P.\ acknowledges support from a Marie Sklodowska-Curie Individual Fellowship within the Horizon 2020 European Union (EU) Framework Programme for Research and Innovation (H2020-MSCA-IF-2014-660113).  Support for this work was also provided by the National Aeronautics and Space Administration (NASA) through an award issued by JPL/Caltech, and through Hubble Fellowship grant HST-HF-51296.01-A awarded by the Space Telescope Science Institute (STScI), which is operated by the Association of Universities for Research in Astronomy, Inc., for NASA, under contract NAS 5-26555.  The Dark Cosmology Centre has been funded by the DNRF.   D.A.P thanks the Aspen Center for Physics for its hospitality and for enabling productive discussion of SLSN models and optical surveys during summer programs in 2014 (``Fast \& Furious'') and 2015 (``The Dynamic Universe''), and particularly acknowledges R.~Chornock's useful presentation on SLSN diversity.  A.V.F. also thanks the Aspen Center for its hospitality during programs in 2014 and 2016 (``Emergence, Evolution, and Effects of Black Holes in the Universe'').   Participation in these programs was supported by National Science Foundation (NSF) Grant No. PHYS-1066293.  D.A.P. acknowledges useful discussions about dwarf-galaxy star-formation histories and surveys with A.~Wetzel and dwarf-galaxy stellar populations with D.~Baade, on SLSN hosts with T.~Chen, on metallicity determinations with M.~Modjaz, and extensive input from G.~Leloudas.  We acknowledge useful feedback on the paper from D.~A.~Kann and D.~Cook.  We thank Y.~Cao, S.~Kulkarni, and M.~Kasliwal for some observations.  We also thank the entire PTF collaboration for their input in building this successful survey, including J.~Bloom for developing and implementing the machine-learning techniques used to identify transients.

A.G.Y. is supported by the EU/FP7 via ERC grant no. 307260, ``The Quantum Universe'' I-Core program by the Israeli Committee for planning and budgeting and the ISF; by Minerva and ISF grants; by the Weizmann-UK ``making connections'' program; and by Kimmel and YeS awards. A.V.F. is grateful for financial assistance from NSF grant AST-1211916, the TABASGO Foundation, Gary and Cynthia Bengier, the Christopher R. Redlich Fund, and NASA/HST grants AR-12850 and AR-14295 from STScI. 

Some of the data presented herein were obtained at the W. M. Keck Observatory, which is operated as a scientific partnership among the California Institute of Technology, the University of California, and NASA. The Observatory was made possible by the generous financial support of the W. M. Keck Foundation.  The authors wish to recognize and acknowledge the very significant cultural role and reverence that the summit of Mauna Kea has always had within the indigenous Hawaiian community.  KAIT and its ongoing operation were made possible by donations from Sun Microsystems, Inc., the Hewlett-Packard Company, AutoScope Corporation, Lick Observatory, the NSF, the University of California, the Sylvia \& Jim Katzman Foundation, and the TABASGO Foundation. Research at Lick Observatory is partially supported by a generous gift from Google.
This paper also includes data based on observations made with the NASA/ESA {\it Hubble Space Telescope} and obtained from the Hubble Legacy Archive, which is a collaboration between STScI/NASA, the Space Telescope European Coordinating Facility (ST-ECF/ESA), and the Canadian Astronomy Data Centre (CADC/NRC/CSA).  This work is based in part on observations made with the  {\it Spitzer Space Telescope}, which is operated by the Jet Propulsion Laboratory, California Institute of Technology, under a contract with NASA.  We also use data products from the Wide-field Infrared Survey Explorer, which is a joint project of the University of California at Los Angeles and the Jet Propulsion Laboratory/California Institute of Technology, funded by NASA.


\bibliographystyle{apj}

\clearpage
\LongTables


\begin{deluxetable*}{llll lll}  
\tabletypesize{\small}
\tablecaption{Photometry of SLSN hosts}
\tablecolumns{7}
\tablehead{
\colhead{PTF ID} &
\colhead{Filter} & 
\colhead{magnitude\tablenotemark{a}} &
\colhead{$F_\nu$\tablenotemark{b}} &
\colhead{date} & 
\colhead{instrument} & 
\colhead{reference}
}
\startdata
09as   & $u$       &  23.01 $\pm$ 0.06 &   2.25 $\pm$ 0.13 & 2016-06-07 & KeckI/LRIS & This work \\
       & $g$       &  22.35 $\pm$ 0.10 &   4.28 $\pm$ 0.41 & 2004-12-21 & SDSS & This work \\
       & $r$       &  21.73 $\pm$ 0.09 &   7.53 $\pm$ 0.65 & 2004-12-21 & SDSS & This work \\
       & $i$       &  21.71 $\pm$ 0.14 &   7.63 $\pm$ 1.05 & 2004-12-21 & SDSS & This work \\
       & $z$       &  21.20 $\pm$ 0.36 &  12.43 $\pm$ 4.89 & 2004-12-21 & SDSS & This work \\

09uy   & $u$       &  24.18 $\pm$ 0.16 &   0.81 $\pm$ 0.13 & 2016-06-07 & KeckI/LRIS & This work \\
       & $B$       &  23.92 $\pm$ 0.05 &   1.20 $\pm$ 0.06 & 2013-04-09 & KeckI/LRIS & This work \\
       & $R$       &  22.53 $\pm$ 0.05 &   3.16 $\pm$ 0.15 & 2013-04-09 & KeckI/LRIS & This work \\
\enddata
\label{tab:photometry}
\tablenotetext{a}{Magnitudes are expressed in the conventional frame for each relevant filter: specifically, SDSS filters ($ugriz$) are reported in the SDSS system, while Johnson-Cousins filters ($BVRI$) are reported in Vega magnitudes.  NIR filters ($JHK_s$) are reported in the 2MASS system.  {\it HST/Spitzer} magnitudes are AB \citep{Oke+1983}; WISE magnitudes are in Vega.  Magnitudes are not corrected for foreground extinction.}
\tablenotetext{b}{Flux densities in $\mu$Jy (calculated from our standard magnitudes via \citealt{Fukugita+1995}), corrected for foreground extinction.}
\end{deluxetable*}

\clearpage

\begin{deluxetable*}{lllll}  
\tabletypesize{\small}
\tablecaption{Log of spectroscopic observations}
\tablecolumns{5}
\tablehead{
\colhead{PTF ID} &
\colhead{Obs.date} &
\colhead{Setup\tablenotemark{a}} &
\colhead{Exposure\tablenotemark{b}} & 
\colhead{PA\tablenotemark{c}}
}
\startdata
09as   & 2016-01-12                  & 400/3400, 400/8500 & 2x900, 2x900     & 235   \\
09uy   & 2014-04-30                  & 400/3400, 400/8500 & 1300, 2x600      &  44   \\
09atu  & 2014-04-29                  & 600/4000, 400/8500 & 1140, 1140       &   5   \\
09cnd  & 2014-05-28                  & 400/3400, 400/8500 & 3x503, 3x500     & 308.5 \\
09cwl  & 2014-04-29                  & 600/5000, 400/8500 & 1000             & 171.5 \\ 
       & 2014-05-28                  & 400/3400, 400/8500 & 3x803, 3x800     & 171.5 \\ 
10bfz  & 2015-04-23                  & 400/3400, 400/8500 & 2x600, 2x600     & 285   \\ 
10bjp  & 2013-12-04                  & 600/4000, 400/8500 & 600+700, 2x600   & 222   \\
\enddata
\label{tab:specobslog}
\tablenotetext{a}{LRIS-B grism and LRIS-R grating, unless another instrument is specified.}
\tablenotetext{b}{Exposure sequence (seconds per exposure).  Comma denotes separate sequences for LRIS-B and LRIS-R.}
\tablenotetext{c}{Sky position angle used for the slit, degrees.  (Zero = North up, with a positive angle indicating rotation counterclockwise on the sky.)}
\tablenotetext{d}{Spectrum contains significant supernova contribution.}
\tablenotetext{e}{Slit centered on the SN and not aligned with the host nucleus.}
\end{deluxetable*}
\clearpage

\begin{deluxetable*}{llll llll l}  
\tabletypesize{\small}
\tablecaption{Photometric properties of PTF SLSN hosts\tablenotemark{a}}
\tablecolumns{9}
\tablehead{
\colhead{PTF ID} &
\colhead{$M_g$} &
\colhead{SFR} &
\colhead{log$_{10}$($M_*$)} &
\colhead{$A_V$} & 
\colhead{Morphology\tablenotemark{b}} \\
\colhead{} &
\colhead{AB mag} &
\colhead{M$_\odot$\,yr$^{-1}$} &
\colhead{M$_\odot$} &
\colhead{mag} &
\colhead{}
}
\startdata
09as   & $-17.72$ & $1.659^{+0.513}_{-1.490}$ & $ 8.27^{+0.92}_{-0.15}$ & $ 1.34^{+0.08}_{-0.95}$ & Compact dwarf \\
09uy   & $-17.99$ & $0.162^{+0.054}_{-0.050}$ & $ 9.24^{+0.11}_{-0.20}$ & $ 0                   $ & Compact dwarf \\
09atu  & $-16.30$ & $0.528^{+0.332}_{-0.245}$ & $ 8.35^{+0.33}_{-0.62}$ & $ 1.53^{+0.19}_{-0.36}$ & Faint dwarf \\
09cnd  & $-17.42$ & $0.481^{+0.203}_{-0.222}$ & $ 8.32^{+0.15}_{-0.18}$ & $ 0.81^{+0.13}_{-0.22}$ & Compact dwarf \\
09cwl  & $-15.30$ & $0.015^{+0.005}_{-0.004}$ & $ 7.85^{+0.23}_{-0.33}$ & $ 0                   $ & Ultra-faint dwarf \\
10bfz  & $-17.48$ & $0.269^{+0.153}_{-0.088}$ & $ 8.65^{+0.11}_{-0.32}$ & $ 0.07^{+0.24}_{-0.07}$ & Compact dwarf \\
10bjp  & $-18.82$ & $0.401^{+0.135}_{-0.120}$ & $ 9.21^{+0.15}_{-0.17}$ & $ 0.01^{+0.05}_{-0.01}$ & Merger \\
10cwr  & $-17.12$ & $0.532^{+0.287}_{-0.248}$ & $ 7.87^{+0.13}_{-0.21}$ & $ 0.81^{+0.15}_{-0.17}$ & Compact dwarf \\
\enddata
\label{tab:hostproperties}
\tablenotetext{a}{Properties derived from SED fitting to the photometry in Table \ref{tab:photometry}.  Uncertainties are derived from the distribution of Monte-Carlo trials plus (for SFR and $M_*$ estimates) an additional systematic uncertainty of 10\%.  An $A_V$ value of 0 with no uncertainty indicates that this parameter was fixed.}
\tablenotetext{b}{Informal morphological properties from visual inspection and from estimates of the size, luminosity, and environment of each host galaxy.}
\end{deluxetable*}

\clearpage

\begin{landscape}
\begin{deluxetable*}{l ccc c ccc cc cc}  
\tabletypesize{\small}
\tablecaption{Emission-Line Fluxes for SLSN hosts\tablenotemark{a}}
\tablecolumns{12}
\tablehead{
\colhead{PTF ID} &
\colhead{${\rm H}\alpha$} &
\colhead{${\rm H}\beta$}  &
\colhead{${\rm H}\gamma$} &
\colhead{[O~II]}  &
\colhead{[O~III]} &
\colhead{[O~III]} &
\colhead{[O~III]} &
\colhead{[N~II]} &
\colhead{[N~II]} &
\colhead{[S~II]} &
\colhead{[S~II]} \\
\colhead{} &
\colhead{6563} &
\colhead{4861} &
\colhead{4340} &
\colhead{3727} &
\colhead{4363} &
\colhead{4959} &
\colhead{5007} &
\colhead{6548} &
\colhead{6584} &
\colhead{6716} &
\colhead{6731} 
}
\startdata
09as                      &$ 62.9\pm 0.6$&$ 18.6\pm 0.7$&$  7.3\pm 0.6$&$ 27.7\pm 0.8$&$  2.1\pm 0.5$&$ 32.4\pm 0.8$&$101.8\pm 1.0$&$  0.7\pm 0.4$&$  2.0\pm 0.4$&$  4.3\pm 0.4$&$  2.8\pm 0.4$\\
09uy                      &$ 11.3\pm 1.6$&$  2.3\pm 0.6$&$< 7.5       $&$  8.7\pm 0.7$&$< 6.3       $&$< 1.8       $&$  4.2\pm 0.6$&$< 2.0       $&$< 5.8       $&$< 6.3       $&$< 3.1       $\\
09atu                     &$< 1.7       $&$< 1.9       $&$< 1.3       $&$< 3.0       $&$< 1.7       $&$< 1.1       $&$< 1.4       $&$< 1.7       $&$< 2.1       $&$< 5.6       $&$< 4.1       $\\
09cnd                     &$ 11.7\pm 0.3$&$  4.7\pm 0.4$&$  1.4\pm 0.4$&$ 10.0\pm 0.6$&$< 1.2       $&$  4.2\pm 0.5$&$ 12.4\pm 1.3$&$< 0.7       $&$< 1.9       $&$  1.8\pm 0.5$&$  1.4\pm 0.4$\\
09cwl                     &$< 1.5       $&$< 1.4       $&$< 2.0       $&$< 1.2       $&$< 4.9       $&$< 1.3       $&$< 1.2       $&$< 2.1       $&$< 3.8       $&$< 1.0       $&$< 0.9       $\\
10bfz                     &$ 80.0\pm 0.8$&$ 26.0\pm 1.5$&$  9.6\pm 0.7$&$ 39.7\pm 1.3$&$  2.6\pm 0.7$&$ 34.9\pm 1.3$&$108.7\pm 1.6$&$< 2.2       $&$  2.1\pm 0.5$&$  7.1\pm 1.3$&$  4.2\pm 1.0$\\
10bjp                     &$ 33.2\pm 0.7$&$ 10.6\pm 0.8$&$< 6.9       $&$ 24.1\pm 0.9$&$< 3.7       $&$ 13.3\pm 0.8$&$ 40.8\pm 1.0$&$< 1.2       $&$< 2.9       $&$  4.0\pm 0.4$&$  2.4\pm 0.4$\\
10cwr                     &$ 33.9\pm 0.7$&$ 13.1\pm 1.2$&$  5.8\pm 1.1$&$ 12.2\pm 1.0$&$  2.9\pm 1.1$&$ 22.8\pm 1.1$&$ 64.0\pm 1.4$&$< 2.0       $&$< 1.6       $&$  1.7\pm 0.5$&$  3.0\pm 0.8$\\
\enddata
\label{tab:linefluxes}
\tablenotetext{a}{In units of $10^{-17}$ erg cm$^{-2}$ s$^{-1}$, corrected for Galactic extinction.  1$\sigma$ statistical uncertainties shown.  These do not include uncertainties associated with the overall flux calibration, which for sources with SN contamination or weak continuum (09atu, 09cwl, 10hgi, 10qwu, 10scc, 10vwg, 12mue, 12mxx) may be significant but will affect all lines proportionally.}
\tablenotetext{b}{For the large disk galaxy near the SLSN position.}
\tablenotetext{c}{From the sum of extractions from a slit through the nucleus of the ``major'' and ``minor'' components of the merger.}
\tablenotetext{d}{From a slit that passes through the ``primary'' galaxy, although not through its nucleus.}
\end{deluxetable*}
\clearpage

\begin{deluxetable*}{lll|lllllllllllll}  
\tabletypesize{\small}
\tablecaption{Emission-Line Diagnostics for SLSN hosts}
\tablecolumns{15}
\tablehead{
\colhead{PTF ID} &
\colhead{SFR$_{{\rm H}\alpha}$\tablenotemark{a}} &
\colhead{$A_{V}$\tablenotemark{b}} &
\multicolumn{12}{c}{12 + log$_{10}$[O/H]} \\ 
\colhead{} &
\colhead{} &
\colhead{} &
\colhead{``Best''\tablenotemark{c}} &
\colhead{$T_e$} &
\colhead{N06 N2} &
\colhead{N06 $R_{23}$} &
\colhead{D02} &
\colhead{PP04 N2} &
\colhead{PP04 O3} &
\colhead{M08} &
\colhead{M13 O3} &
\colhead{M13 N2} &
\colhead{KK04} &
\colhead{KD02} 
}
\startdata
09as  &$0.485^{+0.108}_{-0.107}$&$0.64^{+0.12}_{-0.12}$&$7.98^{+0.05}_{-0.07}$&$7.88^{+0.17}_{-0.12}$&$7.98^{+0.05}_{-0.07}$&$8.07                $&$8.04^{+0.17}_{-0.18}$&$8.08^{+0.03}_{-0.04}$&$8.02^{+0.02}_{-0.03}$&$8.10^{+0.06}_{-0.09}$&$8.06^{+0.02}_{-0.03}$&$8.05^{+0.06}_{-0.06}$&$8.10^{+0.06}_{-0.09}$&$8.10^{+0.06}_{-0.09}$ \\
09uy  &$0.796^{+0.188}_{-0.530}$&$2.00^{+0.07}_{-1.17}$&$8.78^{+0.34}_{-0.34}$&$...                 $&$8.78^{+0.34}_{-0.34}$&$8.07^{+0.59}        $&$8.68^{+0.11}_{-0.19}$&$8.54^{+0.11}_{-0.17}$&$8.47^{+0.05}_{-0.09}$&$8.84^{+0.12}_{-0.19}$&$8.36^{+0.03}_{-0.06}$&$8.47^{+0.06}_{-0.12}$&$8.83^{+0.09}_{-0.16}$&$8.29^{+0.11}_{-0.26}$ \\
09atu &$<0.105                 $&$0                   $&$...                 $&$...                 $&$...                 $&$...                 $&$...                 $&$...                 $&$...                 $&$...                 $&$...                 $&$...                 $&$...                 $&$...                 $ \\
09cnd &$0.115^{+0.023}_{-0.023}$&$0                   $&$8.21^{+0.22}_{-0.71}$&$...                 $&$8.21^{+0.22}_{-0.71}$&$8.63^{+0.04}_{-1.18}$&$8.27^{+0.19}_{-0.31}$&$8.22^{+0.09}_{-0.15}$&$8.23^{+0.06}_{-0.13}$&$8.43^{+0.22}_{-0.32}$&$8.19^{+0.05}_{-0.09}$&$8.21^{+0.10}_{-0.19}$&$8.42^{+0.17}_{-0.38}$&$8.24^{+0.16}_{-0.43}$ \\
09cwl &$<0.061                 $&$0                   $&$...                 $&$...                 $&$...                 $&$...                 $&$...                 $&$...                 $&$...                 $&$...                 $&$...                 $&$...                 $&$...                 $&$...                 $ \\
\enddata
\label{tab:linediagnostics}
\tablenotetext{a}{Dust-corrected H$\alpha$ SFR, in $M_\odot$\,yr$^{-1}$.}
\tablenotetext{b}{Balmer-decrement dust extinction, in mag.}
\tablenotetext{c}{The ``best'' metallicity (the value presented in plots and elsewhere in the discussion) is the $T_e$ metallicity when a high-significance detection of the [O~III] $\lambda{4363}$ auroral line is available, else the N06 (N2H$\alpha$) metallicity if [N~II] is detected.  The mass-metallicity relation is used if neither of these lines is detected.}
\end{deluxetable*}

\clearpage
\end{landscape}

\end{document}